\PassOptionsToPackage{frozencache}{minted}
\def\arxiv{}
\ifdefined\arxiv
  \documentclass[sigplan,10pt,nonacm]{acmart}
\else
  \documentclass[sigplan,10pt,anonymous,review]{acmart}
\fi

\newif\ifarxiv
\ifdefined\arxiv
  \arxivtrue
  \newcommand{\dcRepo}{\url{https://github.com/PierreSenellart/descriptive-complexity}}
\fi

\ifarxiv\else
\acmConference[CPP '27]{ACM SIGPLAN International Conference on Certified
  Programs and Proofs}{January 11--12, 2027}{Mexico City, Mexico}
\acmYear{2027}
\fi

\usepackage{array}
\newcolumntype{P}[1]{>{\raggedright\arraybackslash}p{#1\textwidth}}

\usepackage{tikz}
\usetikzlibrary{shapes,arrows.meta}
\usepackage[shell,tikz]{dot2texi}

\definecolor{karpfill}{HTML}{E8EEF7}
\tikzset{
  rounded/.style={rounded corners=2pt},
  filled/.style={fill=white},
  karp/.style={fill=karpfill},
  legend/.style={font=\fontsize{22}{25}\selectfont},
}

\makeatletter
\newcommand{\dotfigure}[2][]{%
  \begingroup
    \let\dtt@saved@definecolor\definecolor
    \renewcommand{\definecolor}[3]{%
      \dtt@saved@definecolor{##1}{##2}{##3}\@ifnextchar;{\@gobble}{}}%
    \setkeys{dtt}{dot,tikz,#1,file=#2}%
    \dottotexgraphicsinclude
  \endgroup}
\makeatother

\usepackage{apxproof}
\usepackage{booktabs}
\usepackage{enumitem}
\usepackage{minted}
\usepackage{newunicodechar}
\usepackage{stmaryrd}
\usepackage{xspace}
\xspaceaddexceptions{(}

\setminted{
  fontsize=\small,
  breaklines=true,
  breakanywhere=false,
  autogobble=true,
  xleftmargin=1em,
  numbersep=0.5em,
  style=tango
}
\setminted[lean4]{escapeinside=@@, ignorelexererrors=true}

\newminted[leancode]{lean4}{}
\newmintedfile[leanfile]{lean4}{}
\newmintinline[lean]{lean4}{}

\makeatletter
\begingroup
\catcode`\_=12\relax
\gdef\decl@us{_}
\endgroup
\begingroup
\gdef\decl@dot{.}
\gdef\decl@usplit#1_#2\@nil{%
  \decl@dsplit#1.\@nil
  \def\decl@urest{#2}%
  \ifx\decl@urest\@empty\else
    \decl@us\discretionary{}{}{}%
    \decl@usplit#2\@nil
  \fi}
\gdef\decl@dsplit#1.#2\@nil{%
  #1%
  \def\decl@drest{#2}%
  \ifx\decl@drest\@empty\else
    \decl@dot\discretionary{}{}{}%
    \decl@dsplit#2\@nil
  \fi}
\gdef\decl@arm#1{\decl@usplit#1_\@nil}
\endgroup
\DeclareRobustCommand{\decl}[1]{\texttt{\decl@arm{#1}}}
\makeatother

\newcommand{\nb}[1]{\hbox to 0pt{\textcolor{red}{!}\hss}\marginpar{\raggedright\footnotesize\textcolor{red}{#1}}}
\newcommand{\na}[1]{\hbox to 0pt{\textcolor{blue}{!}\hss}\marginpar{\raggedright\footnotesize\textcolor{blue}{A: #1}}}

\newunicodechar{≤}{\ensuremath{\leq}}
\newunicodechar{≃}{\ensuremath{\simeq}}
\newunicodechar{≡}{\ensuremath{\equiv}}
\newunicodechar{≐}{\ensuremath{\doteq}}
\newunicodechar{⋯}{\ensuremath{\cdots}}
\newunicodechar{ᶜ}{\ensuremath{^{\mathrm{c}}}}
\newunicodechar{⟨}{\ensuremath{\langle}}
\newunicodechar{⟩}{\ensuremath{\rangle}}
\newunicodechar{∘}{\ensuremath{\circ}}
\newunicodechar{ι}{\ensuremath{\iota}}
\newunicodechar{ᶠ}{\ensuremath{^{\mathrm{f}}}}
\newunicodechar{ʳ}{\ensuremath{^{\mathrm{r}}}}
\newunicodechar{≥}{\ensuremath{\geq}}
\newunicodechar{≠}{\ensuremath{\neq}}
\newunicodechar{∈}{\ensuremath{\in}}
\newunicodechar{∉}{\ensuremath{\notin}}
\newunicodechar{⊆}{\ensuremath{\subseteq}}
\newunicodechar{∀}{\ensuremath{\forall}}
\newunicodechar{∃}{\ensuremath{\exists}}
\newunicodechar{¬}{\ensuremath{\lnot}}
\newunicodechar{∧}{\ensuremath{\land}}
\newunicodechar{∨}{\ensuremath{\lor}}
\newunicodechar{→}{\ensuremath{\rightarrow}}
\newunicodechar{↔}{\ensuremath{\leftrightarrow}}
\newunicodechar{⇒}{\ensuremath{\Rightarrow}}
\newunicodechar{↦}{\ensuremath{\mapsto}}
\newunicodechar{α}{\ensuremath{\alpha}}
\newunicodechar{β}{\ensuremath{\beta}}
\newunicodechar{γ}{\ensuremath{\gamma}}
\newunicodechar{σ}{\ensuremath{\sigma}}
\newunicodechar{φ}{\ensuremath{\varphi}}
\newunicodechar{ψ}{\ensuremath{\psi}}
\newunicodechar{Σ}{\ensuremath{\Sigma}}
\newunicodechar{Π}{\ensuremath{\Pi}}
\newunicodechar{ℕ}{\ensuremath{\mathbb{N}}}
\newunicodechar{ᵒ}{\ensuremath{^{\mathrm{o}}}}
\newunicodechar{ᵖ}{\ensuremath{^{\mathrm{p}}}}
\newunicodechar{ₖ}{\ensuremath{_{k}}}
\newunicodechar{₀}{\ensuremath{_{0}}}
\newunicodechar{₁}{\ensuremath{_{1}}}
\newunicodechar{₂}{\ensuremath{_{2}}}
\newunicodechar{₃}{\ensuremath{_{3}}}
\newunicodechar{₄}{\ensuremath{_{4}}}
\newunicodechar{⊓}{\ensuremath{\sqcap}}
\newunicodechar{⊥}{\ensuremath{\bot}}
\newunicodechar{⊤}{\ensuremath{\top}}
\newunicodechar{∼}{\ensuremath{\sim}}
\newunicodechar{×}{\ensuremath{\times}}
\newunicodechar{⊕}{\ensuremath{\oplus}}

\newcommand{\class}[1]{\ensuremath{\mathsf{#1}}\xspace}
\newcommand{\Ptime}{\class{PTIME}}
\newcommand{\NP}{\class{NP}}
\newcommand{\coNP}{\class{coNP}}
\newcommand{\Logspace}{\class{LOGSPACE}}
\newcommand{\NL}{\class{NL}}
\newcommand{\coNL}{\class{coNL}}
\newcommand{\ACz}{\ensuremath{\class{AC}^{0}}}
\newcommand{\PSPACE}{\class{PSPACE}}
\newcommand{\EXPTIME}{\class{EXPTIME}}
\newcommand{\NEXPTIME}{\class{NEXPTIME}}
\newcommand{\EXPSPACE}{\class{EXPSPACE}}
\newcommand{\coPSPACE}{\class{coPSPACE}}
\newcommand{\SigmaP}[1]{\ensuremath{\Sigma^{\mathrm{p}}_{#1}}}
\newcommand{\PiP}[1]{\ensuremath{\Pi^{\mathrm{p}}_{#1}}}
\newcommand{\PH}{\class{PH}}
\newcommand{\RE}{\class{RE}}
\newcommand{\DP}{\class{DP}}
\newcommand{\GI}{\class{GI}}

\newcommand{\FO}{\ensuremath{\mathrm{FO}}\xspace}
\newcommand{\SO}{\ensuremath{\mathrm{SO}}\xspace}
\newcommand{\ESO}{\ensuremath{\exists\SO}\xspace}
\newcommand{\ESOnew}{\ensuremath{\ESO_{\mathrm{new}}}\xspace}
\newcommand{\ESOnewbd}[1]{\ensuremath{\ESO_{\mathrm{new}}[#1]}\xspace}
\newcommand{\ASO}{\ensuremath{\forall\SO}\xspace}
\newcommand{\TC}{\ensuremath{\mathrm{TC}}\xspace}
\newcommand{\DTC}{\ensuremath{\mathrm{DTC}}\xspace}
\newcommand{\LFP}{\ensuremath{\mathrm{LFP}}\xspace}
\newcommand{\IFP}{\ensuremath{\mathrm{IFP}}\xspace}
\newcommand{\PFP}{\ensuremath{\mathrm{PFP}}\xspace}
\newcommand{\expof}[1]{\ensuremath{\mathsf{#1}^{\mathrm{exp}}}\xspace}

\newcommand{\fored}{\ensuremath{\preccurlyeq_{\mathrm{fo}}}\xspace}
\newcommand{\foredord}{\ensuremath{\preccurlyeq_{\mathrm{fo}}^{\leq}}\xspace}
\newcommand{\foredrel}{\ensuremath{\preccurlyeq_{\mathrm{rfo}}^{\leq}}\xspace}

\newcommand{\prob}[1]{\textsc{#1}}

\newcommand{\imp}{\rightarrow}
\newcommand{\p}{\varphi}
\newcommand{\mdl}{\models}
\newcommand{\sbs}{\subseteq}
\newcommand{\old}{\mathsf{old}}
\newcommand{\Holds}{\ensuremath{\mathtt{Holds}}\xspace}

\newcommand{\Cmc}{\ensuremath{\mathcal{C}}\xspace}

\newcommand{\Lmc}{\ensuremath{\mathcal{L}}\xspace}

\newcommand{\aol}{\ensuremath{\overline{a}}\xspace}

\newcommand{\tol}{\ensuremath{\overline{t}}\xspace}

\newcommand{\xol}{\ensuremath{\overline{x}}\xspace}
\newcommand{\yol}{\ensuremath{\overline{y}}\xspace}

\newcommand{\Att}{\ensuremath{\mathtt{A}}\xspace}

\newcommand{\Ltt}{\ensuremath{\mathtt{L}}\xspace}

\newcommand{\Ptt}{\ensuremath{\mathtt{P}}\xspace}
\newcommand{\Qtt}{\ensuremath{\mathtt{Q}}\xspace}

\newcommand{\Wtt}{\ensuremath{\mathtt{W}}\xspace}

\newcommand{\arxivCategories}{7}
\newcommand{\arxivHardness}{745}
\newcommand{\arxivHardnessPct}{15.5}
\newcommand{\arxivPapers}{4\,795}
\newcommand{\arxivProves}{475}
\newcommand{\arxivProvesPct}{9.9}
\newcommand{\arxivProvesPctMax}{24.8}
\newcommand{\arxivProvesPctMin}{6.0}
\newcommand{\balbachExtLines}{16\,838}

\newcommand{\balbachModuleLines}{50\,509}
\newcommand{\balbachModules}{18}
\newcommand{\balbachOwnLines}{43\,946}

\newcommand{\clSatExtLines}{25\,796}
\newcommand{\clSatModuleLines}{38\,017}
\newcommand{\clSatModules}{56}
\newcommand{\clSatOwnLines}{29\,693}
\newcommand{\dcBothExtLines}{14\,996}
\newcommand{\dcBothModuleLines}{9\,236}
\newcommand{\dcBothModules}{32}
\newcommand{\dcBothOwnLines}{6\,262}
\newcommand{\dcCompleteClasses}{14}
\newcommand{\dcCompleteThms}{73}
\newcommand{\dcFiles}{726}

\newcommand{\gkExtLines}{25\,762}

\newcommand{\gkModuleLines}{24\,322}
\newcommand{\gkModules}{72}
\newcommand{\gkOwnLines}{19\,658}

\newcommand{\completeproblemrows}{%
  \Logspace{} & 2 \\
  \NL{} & 3 \\
  \Ptime{} & 4 \\
  \NP{} & 38 \\
  \coNP{} & 3 \\
  \DP{} & 1 \\
  \SigmaP{k} & 2 \\
  \PiP{k} & 2 \\
  \PH{} & -- \\
  \PSPACE{} & 5 \\
  \EXPTIME{} & 1 \\
  \NEXPTIME{} & 2 \\
  \EXPSPACE{} & 3 \\
  \RE{} & 4 \\
  \midrule
  \GI{} & 3 \\%
}
\newcommand{\dcLinesK}{212k}

\newcommand{\mlEXPTIMETotal}{12\,641}
\newcommand{\mlEXPTIMEMarginal}{8\,471}

\newcommand{\mlNEXPTIMETotal}{40\,832}

\newcommand{\mlEXPSPACETotal}{47\,007}

\newcommand{\mlEXPSPACEHard}{46\,161}

\newcommand{\mlNPaltMarginal}{442}

\newcommand{\mlPSPACEnondetMarginal}{124}

\newcommand{\mlEXPSPACEdetMarginal}{55}

\newcommand{\mlEXPTIMETotalOver}{4\,746}

\newcommand{\machinelevelrows}{%
  \Logspace & \decl{HeadAutomaton} (det.)\textsuperscript{$\ast$} & 22 & -- & -- & 5\,035 & 4\,024 & \prob{det-reach} \\
  \NL & \decl{HeadAutomaton}\textsuperscript{$\ast$} & 31 & -- & -- & 6\,987 & 2\,279 & \prob{reach} \\
  \Ptime & \decl{DTMAccept} & 37 & 4\,707 & 4\,731 & 7\,895 & 3\,897 & \prob{cvp} \\
  \NP & \decl{NTMAccept} & 32 & 2\,485 & 5\,640 & 6\,268 & 2\,949 & \prob{sat} \\
  \NP & \decl{ATMAccept} (1 block)\textsuperscript{$\dagger$} & 35 & 2\,848 & 5\,978 & 6\,706 & 442 & \decl{NTMAccept} \\
  \coNP & \decl{ATMAccept} & 57 & -- & -- & 10\,908 & 7\,776 & \prob{taut} \\
  \SigmaP{k} & \decl{ATMAccept} & 57 & 4\,411 & 8\,586 & 10\,871 & 6\,522 & \prob{qbf}\(_k\) \\
  \PiP{k} & \decl{ATMAccept} & 57 & 4\,444 & 8\,620 & 10\,905 & 6\,522 & \prob{qbf}\(^\forall_k\) \\
  \PSPACE & \decl{DTMAcceptSpace} & 51 & 1\,859 & 9\,302 & 9\,807 & 3\,435 & \prob{qsat} \\
  \PSPACE & \decl{NTMAcceptSpace}\textsuperscript{$\dagger$} & 52 & 1\,750 & 9\,462 & 9\,888 & 124 & \decl{DTMAcceptSpace} \\
  \EXPTIME & \decl{ATMAcceptSpace} & 87 & 5\,151 & 11\,639 & 12\,641 & 8\,471 & \decl{DTMAccept} \\
  \NEXPTIME & \decl{WideRegAccept} & 209 & 3\,464 & 39\,584 & 40\,832 & 38\,162 & \decl{NTMAccept} \\
  \EXPSPACE & \decl{WideAcceptSpace} & 207 & 11\,759 & 46\,161 & 47\,007 & 37\,243 & \decl{NTMAcceptSpace} \\
  \EXPSPACE & \decl{DWideAcceptSpace}\textsuperscript{$\dagger$} & 206 & 11\,791 & 46\,047 & 46\,924 & 55 & \decl{WideAcceptSpace} \\
  \RE & \decl{HALT} & 59 & 3\,170 & 14\,253 & 15\,614 & 8\,890 & \prob{finsat} \\
}

\newcommand{\problemCount}{42}
\newcommand{\problemMachineThms}{12}

\newcommand{\problemMedian}{869}

\newcommand{\problemQOne}{521}
\newcommand{\problemQThree}{1\,192}
\newcommand{\problemSubjects}{68}
\newcommand{\problemTheorems}{61}

\newcommand{\problemUnderTwoThousand}{35}

\newcommand{\leanVersion}{v4.33.0}
\newcommand{\mathlibPin}{v4.33.0}
\newcommand{\problemcostrows}{%
  \decl{Taut} & \coNP & 1 & 147 & \decl{ConjunctiveQueries} & \NP & 3 & 878 \\%
  \decl{ThreeDnfTaut} & \coNP & 2 & 159 & \decl{Cvp} & \Ptime & 1 & 881 \\%
  \decl{NaeThreeSat} & \NP & 1 & 256 & \decl{TwoSat} & \NL & 1 & 922 \\%
  \decl{NaeSat} & \NP & 1 & 270 & \decl{MaxCut} & \NP & 1 & 924 \\%
  \decl{DigraphIso} & \GI & 1 & 295 & \decl{HornSat} & \Ptime & 1 & 944 \\%
  \decl{Reachability} & \NL & 2 & 356 & \decl{GraphCrawling} & \NP & 2 & 964 \\%
  \decl{WideCorridor} & \EXPSPACE & 1 & 401 & \decl{Qbf} & \PiP{k}, \SigmaP{k} & 2 & 968 \\%
  \decl{SatUnsat} & \DP & 1 & 423 & \decl{Knapsack} & \NP & 1 & 973 \\%
  \decl{DominatingSet} & \NP & 1 & 434 & \decl{ThreeDimMatching} & \NP & 1 & 1\,025 \\%
  \decl{SubgraphIso} & \NP & 1 & 474 & \decl{GraphIso} & \GI & 1 & 1\,036 \\%
  \decl{ZeroOneIP} & \NP & 1 & 521 & \decl{Steiner} & \NP & 2 & 1\,192 \\%
  \decl{Coloring} & \NP & 3 & 558 & \decl{SuccinctReach} & \PSPACE & 1 & 1\,363 \\%
  \decl{ThreeColorability} & \NP & 1 & 568 & \decl{Partition} & \NP & 1 & 1\,482 \\%
  \decl{Game} & \Ptime & 1 & 607 & \decl{Sat} & \NP & 1 & 1\,730 \\%
  \decl{ReachabilityDet} & \Logspace & 2 & 617 & \decl{Hamilton} & \NP & 2 & 2\,032 \\%
  \decl{ThreeSat} & \NP & 1 & 622 & \decl{JobSequencing} & \NP & 1 & 2\,164 \\%
  \decl{Feedback} & \NP & 2 & 677 & \decl{WideTiling} & \NEXPTIME & 1 & 2\,516 \\%
  \decl{DagIso} & \GI & 1 & 711 & \decl{CodeHalt} & \RE & 1 & 3\,051 \\%
  \decl{CliqueFamily} & \NP & 3 & 750 & \decl{Qsat} & \PSPACE & 2 & 3\,609 \\%
  \decl{SetFamily} & \NP & 5 & 836 & \decl{Pcp} & \RE & 1 & 4\,170 \\%
  \decl{OneInSat} & \NP & 1 & 861 & \decl{FinSat} & \RE & 1 & 5\,038 \\
}

\begin{document}

\title{Descriptive Complexity in Lean:
  Completeness~by~\mbox{First-Order}~Reductions}

\author{Pierre Senellart}
\orcid{0000-0002-7909-5369}
\affiliation{%
  \institution{DI ENS, ENS, PSL University, CNRS, Inria}
  \city{Paris}
  \country{France}
}
\email{pierre@senellart.com}
\author{Anton Gnatenko}
\orcid{0000-0003-1499-2090}
\affiliation{%
  \institution{DI ENS, ENS, PSL University, CNRS, Inria}
  \city{Paris}
  \country{France}
}
\email{anton.gnatenko@ens.psl.eu}

\begin{abstract}
  We show that descriptive complexity can serve as a foundation for
  formalizing computational complexity results in a proof assistant, by
  constructing a Lean library centered around the following concepts:
  decision problems are isomorphism-invariant predicates on finite
  structures;
  complexity classes are defined by their logical characterization;
  membership is shown by definability witnesses; hardness is shown by
  first-order reductions from a known hard problem. We also establish
  bridges to traditional machine models such as (non)deterministic Turing
  machines. The library proves \dcCompleteThms{} completeness results, on
  \problemSubjects{} problems or problem families, over
  \dcCompleteClasses{} different classes; relations between the classes
  established inside the logic and not by machine simulation, among them
  $\NL = \coNL$ and the Abiteboul--Vianu theorem; and unconditional lower
  bounds, among them $\FO({\leq}) \subsetneq \FO({\leq}, \TC)$ and the failure of
  order-free \FO(\IFP) to capture \Ptime{}.
\end{abstract}

\ccsdesc[500]{Theory of computation~Complexity classes}
\ccsdesc[500]{Theory of computation~Problems, reductions and completeness}
\ccsdesc[500]{Mathematics of computing~Mathematical software}
\ccsdesc[300]{Theory of computation~Finite Model Theory}
\ccsdesc[300]{Theory of computation~Logic and verification}

\keywords{descriptive complexity, computational complexity, first-order
  reductions, NP-completeness, complexity classes, finite model theory, Lean,
  Mathlib}

\maketitle

\section{Introduction}
\label{sec:intro}

\begin{toappendix}
\section{How Often the Literature Proves a Hardness Result}
\label{sec:appendix:survey}

We explain how the number \S\ref{sec:intro} opens with was measured. We took
every arXiv submission of 2025 whose \emph{primary} category is one of the
\arxivCategories{} theoretical computer science categories, \arxivPapers{}
records. For each we checked whether its title or abstract announces a
hardness, completeness or undecidability result, and whether it announces one
as the authors' own.
\ifarxiv\else
The corpus ships frozen with the supplementary material, so the classification
can be rerun offline.
\fi

We measure what a paper announces in title or abstract only, so
\arxivProves{} is a lower bound. Mentioning is not claiming either:
\arxivHardness{} papers (\arxivHardnessPct\%) mention such a result, most
of them to justify a heuristic, while \arxivProves{} (\arxivProvesPct\%)
claim one of their own. We use simple English patterns: to count as
claiming, a paper must put a claim verb and a hardness term in one
sentence, the verb first.

We need to pay attention to three potential pitfalls. Abstracts carry
markup, so a pattern written against \textsf{NP}-complete misses the same words
set as mathematics, and we strip the markup first. The OAI-PMH interface,
recommended for bulk harvesting, keys its windows on a datestamp that moves to
the latest revision, so a December paper revised in January is assigned
the wrong year; we use the query interface, which filters on the submission
date. And ``complete for'' is overloaded, as in ``sound and complete for
bounded postconditions'', so we require a complexity class to follow it.

We hand-checked the classifier against samples. Of 60 sampled positives, 58
are correct. The two failures are a paper writing ``given the \emph{known}
\NP-hardness'' of its problem, and a claimed (obviously bogus) \Ptime{} versus
\NP{} separation in which the \NP-completeness of \prob{sat} is background,
not the result. Precision is therefore near 96\%. Of 40 sampled papers that
mention a hardness result without being credited with proving one, three were
misses, all gaps in the list of claim verbs; adding those verbs caught six
papers across the corpus and moved the count from 469 to \arxivProves{}. Of 42
further sampled papers with a weaker cue or none, none were misses. The two
kinds of error are of comparable size and opposite in sign, so \arxivProves{}
is accurate to within a few percent.

We checked the choice of arXiv categories for theoretical computer
science. Leaving out a theoretical one would
understate the count, and taking in an applied one would dilute it. The seven
differ widely among themselves, from \arxivProvesPctMin\% (cs.LO) to
\arxivProvesPctMax\% (cs.CC), and every plausible neighbor sits an order of
magnitude below: cs.DB at 1.2\%, cs.SC at 1.1\%, cs.PL at 0.7\%, and cs.DC,
cs.IT, cs.CR and cs.MS below those. Over all of cs.* the proportion would be
0.5\%, due to the archive current dominance by machine learning papers.

Two caveats remain. First, arXiv is not the literature: it mixes preprints with
published work, and authors choose their own primary category. Second,
\S\ref{sec:intro} says such results are almost always established by
reduction. We checked that only on a sample: of 40 counted papers read in
full, 35 use explicit reduction language.

\end{toappendix}

Of the \arxivPapers{} papers submitted to arXiv in 2025 whose primary
category is one of the seven theoretical computer science
categories\footnote{cs.CC, cs.DM, cs.FL, cs.DS, cs.GT, cs.CG, and
cs.LO.}, \arxivProves{} (or \arxivProvesPct\%) announce, in their title
or abstract, a new computational complexity hardness or completeness
result (see Appendix~\ref{sec:appendix:survey}).
Such results are almost always established by reduction from a known hard
problem, and are essentially never formalized using a proof assistant.

Indeed, proof assistant libraries provide very limited support for
computational complexity. Lean's Mathlib~\cite{mathlib2020} defines no
computational complexity class, and no reduction carrying a resource bound
(\S\ref{sec:landscape}).
The few examples of computational complexity formalization each stop at a
handful of completeness
theorems~\cite{gaeher2021cooklevin,balbach2023cooklevin,complexitylib}, or at
reductions whose hardness root is left admitted~\cite{polyreductions}
(\S\ref{sec:landscape}).

We claim the main reason is that standard computational complexity proofs are
anchored to a machine model under resource constraints, and that such models
resist mechanization. This argument was made by the researchers who attempted
to prove otherwise. \citet{forster2020turing} built a framework for
programming and verifying multi-tape Turing machines in Rocq (formerly
Coq)~\cite{rocq}, and used it to verify a universal machine and a multi-tape
to single-tape compiler. Their conclusion, after some nineteen thousand lines,
was nonetheless negative, and blamed not their tooling but the model: ``Turing
machines as model of computation are inherently infeasible for the
formalisation of any computability or complexity theoretic result''.
\citet{gaeher2021cooklevin} obtained the first complexity-theoretic result
mechanized against a concrete computational model, the Cook--Levin theorem, by
giving up Turing machines for the call-by-value~$\lambda$-calculus; even then
machines survive as an intermediate problem in the reduction chain, and the
resource analysis remains ``a major overhead over just verifying the
functional correctness of reductions''. \mbox{\citet{balbach2023cooklevin}}
pays that cost in full, formalizing Cook--Levin over multi-tape machines in
Isabelle/HOL~\cite{nipkow2002isabelle}.

Once the \NP-hardness of \prob{sat} was
proved~\cite{gaeher2021cooklevin,balbach2023cooklevin}, further completeness
results might have followed easily. They did not: the four developments we
compare in Table~\ref{tab:comparison} hold seven completeness theorems between
them, and the three of~\citet{gaeher2021cooklevin} all sit in the module that
proves Cook--Levin itself. Each further proof again depends on a computational
model, needing a proof that its reduction runs under bounded resources.
A solution to this was named by the authors of
\cite{gaeher2021cooklevin} themselves, who close by asking how
``characterisations of \Ptime{} and \NP{} that are independent from a
computational model, like via Fagin's theorem'', relate to their own.
This paper takes that route and demonstrates that descriptive complexity
is fruitful to formalize complexity results and proofs in a proof
assistant, namely Lean~\cite{moura2021lean4}.

\medskip

\emph{Descriptive complexity}~\cite{immerman1999descriptive} is the
subfield of finite model theory~\cite{libkin2004elements} concerned with
logical characterizations of complexity classes. Its best known result is
Fagin's theorem~\cite{fagin1974generalized}: the class \NP{} \emph{is} the
set of decision problems definable in the existential fragment of
second-order logic (\ESO). Such characterizations make it possible to prove
complexity results while sidestepping the machine model and its resource
constraints entirely. A complexity class is defined by its logical
characterization; a problem is shown to belong to it by exhibiting a
sentence of the corresponding logic that defines it; and hardness is
established through \emph{first-order} (\FO) reductions, building an
instance of one problem from an instance of another through a family of
first-order formulas.
\FO{} reductions are strictly weaker than
Karp reductions, so hardness proved
through them is the stronger statement. And they need only first-order
logic, which Mathlib already provides.

Our contributions are as follows:
\begin{enumerate}[label=(\roman*),nosep]
  \item A framework for descriptive complexity in Lean
    (\S\ref{sec:framework}): decision problems as isomorphism-invariant
    predicates on finite structures, reductions as first-order interpretations,
    and an encoding mechanism for converting user-defined instances into this form.

  \item Complexity classes defined by their logical characterizations
    (\S\ref{sec:classes}), from \Logspace{} to \RE{} and along the
    polynomial hierarchy, closed under \FO{} reductions by construction,
    together with properties proved inside the logic: $\NL = \coNL$ and $\PSPACE = \coPSPACE$, the fixed-point
    characterizations of \Ptime{} and \PSPACE{}, the Abiteboul--Vianu
    theorem~\cite{abiteboul1995computing,abiteboul1991generic}, and, on
    Ehrenfeucht--Fra{\"i}ss{\'e} games built for the purpose, the
    unconditional lower bounds $\FO({\leq}) \subsetneq \FO({\leq}, \TC)$ and
    the failure of order-free \FO(\IFP) to capture \Ptime{}.

  \item Machine bridges (\S\ref{sec:machines}) that validate the approach
    outside the logic: e.g., we show that \NP{}, defined as $\ESO$, is exactly
    the class of problems reducible to the acceptance problem for
    nondeterministic machines. The machine, its tape contents and its
    computations are encoded in a finite structure, with the count of the
    universe elements bounding the running time. We prove Cook--Levin in both
    its machine-free and its textbook form.

  \item \dcCompleteThms{} completeness theorems across
    \dcCompleteClasses{} classes on concrete problems
    (\S\ref{sec:problems}), by first-order reductions.
\end{enumerate}

\noindent
We then compare to existing mechanized complexity theory~(\S\ref{sec:related})
and discuss scope and limitations
(\S\ref{sec:design}).
The library is \dcLinesK{} lines of Lean across \dcFiles{} modules,
using Mathlib~\mathlibPin{} on Lean~\leanVersion{},
\ifarxiv
available at \dcRepo{}.
\else
provided as supplementary material, with the scripts that compute every
measured number here.
\fi

\section{Background}
\label{sec:background}

In his PhD thesis, Ronald Fagin~\citep{fagin1973thesis,fagin1974generalized}
proved that existential second-order logic \emph{captures}~\NP{}:
a property of finite structures is decidable in nondeterministic polynomial
time if and only if it is expressible by a sentence in this logic.
This finding kicked off the study of computational complexity through logic and
finite model theory. We recall the relevant notions and results below and refer
the reader to~\mbox{\citet{immerman1999descriptive}},
\mbox{\citet{ebbinghaus1995finite}} and~\mbox{\citet{libkin2004elements}} for
more details.

\subsection{Finite structures and logics}
\label{sec:background:fsl}

Fix a \emph{relational vocabulary} $\Ltt$, or \emph{vocabulary} for short,
that is, a finite set of relation symbols, each with a prescribed arity. An
\emph{$\Ltt$-structure} $\Att$ consists of a set~$\Delta_\Att$, called its
\emph{domain} or \emph{universe}, and, for each $k$-ary relation symbol~$R \in
\Ltt$, an interpretation $R^\Att \subseteq \Delta_\Att^k$. For example, a
directed graph~$G$ can be seen as an~$\{E\}$-structure over the universe of
its nodes, with~$(a, b) \in E^{\,G}$ whenever there is an edge from $a$ to
$b$.

First-order logic (\FO{}) formulas are built from atoms of the forms
$R(x_1,\ldots,x_k)$ and $x_i=x_j$ using Boolean connectives and quantification
over the elements of the universe. A \emph{sentence} is a formula without free
variables. For example, $\exists x\, \exists y\, (x \neq y \land E(x,y))$
expresses that a graph has at least one edge. We write $\Att \mdl \varphi$ to
say that $\varphi$ is \emph{satisfied} in~$\Att$.

Complexity is measured over \emph{finite} structures, i.e., those with finite
universes, and every statement below quantifies over those only. Satisfaction,
on the other hand, is defined at any cardinality: finiteness is a hypothesis
of the statements, not part of the semantics~(\S\ref{sec:framework:problems}).

We use several extensions of \FO by \emph{closure and fixpoint operators}.
The transitive-closure operator \TC forms the transitive closure of a binary
relation defined by a formula. For example, $\TC[E](x, y)$ says ``there is a path from $x$ to~$y$''.
The deterministic operator \DTC{} is restricted to formulas
that define functional binary relations.

Now, suppose that a formula $\p(\xol)$ uses a fresh relation symbol
$R\notin \Ltt$ of arity $|\xol|$, and that $R$ occurs in $\p$ only
positively. It induces a monotone operator $F_\p$ on $|\xol|$-ary relations over
$\Delta_\Att$, defined by
\[
  F_\p(X) = \{\,\aol\in\Delta_\Att^{|\xol|} \mid (\Att \text{ with } R^\Att = X)\mdl\p(\aol)\,\}.
\]
Then $\LFP[\p](\xol)$ holds if and only if $\xol$ belongs to the least fixed point of $F_\p$.
For example, graph reachability is equivalently defined by the formula
$\LFP[x = y \vee \exists z\, (R(x,z)\land E(z,y))](x,y)$.
Two variants appear below. The \emph{inflationary} fixed point, \IFP{}, is
that of the inflationary operator $F'_\p(X) = X \cup F_\p(X)$, so it converges
without asking $R$ to occur positively; the \emph{partial} fixed point \PFP{}
iterates $F_\p$ from the empty relation under no restriction at all, and
returns the limit when the iteration stabilizes and the empty relation when it
does not.

We also use second-order logic, which admits \emph{relational variables} of various arities.
Over a graph, for example, a
unary relational variable represents a set of vertices, while a ternary relation
variable stands for a set of triples. An \ESO{} sentence has
only existential second-order quantifiers. It can, for example, say that the graph is bipartite:
\[
  \exists X\, \forall x\, \forall y\,
  \bigl(E(x, y) \rightarrow (X(x) \leftrightarrow \neg X(y))\bigr).
\]
Its universal counterpart, \ASO{}, permits only universal second-order
quantifiers.
More generally, the second-order alternation hierarchy consists of the
fragments~$\Sigma^1_k$ and $\Pi^1_k$, for $k \geq 1$. A sentence in
$\Sigma^1_k$ has a prefix with at most $k$ alternating blocks of second-order
quantifiers, beginning with an existential block; $\Pi^1_k$ is defined
dually, beginning with a universal block. Thus, $\Sigma^1_1$ is \ESO{} and
$\Pi^1_1$ is \ASO{}.

The Horn and Krom fragments require the first-order matrix of an \ESO{}
sentence to be a universally quantified conjunction of clauses, disjunctions
of literals, restricted in their second-order literals only, with first-order
atoms occurring freely as a \emph{guard}~$\gamma$ on the clause. A \emph{Horn}
clause has at most one positive second-order literal, so it reads as a rule
$\gamma \wedge X_1(\xol_1) \wedge \dots \wedge X_n(\xol_n) \imp X(\yol)$ over
the quantified relation variables, or as the same with~$\bot$ on the right; a
\emph{Krom} clause has at most two second-order literals. The respective
fragments are denoted \SO-Horn and \SO-Krom, and over ordered structures,
where the guards may use~$\leq$, $\SO\text{-Horn}({\leq})$ and
$\SO\text{-Krom}({\leq})$.

Finally, a logic can be extended with \emph{value invention}, in the style of
the object-creating query languages
of~\citet[ch.~18]{abiteboul1995foundations}. An \ESOnew{} sentence is an
\ESO{} sentence whose relation variables range over the universe of~$\Att$
extended with finitely many fresh elements, related to nothing, with a unary
predicate $\old \notin \Ltt$ marking the original ones. How many may be
invented is itself a resource: holding them to the number of $d$-tuples of the
instance gives \ESOnewbd{d}, holding them to exponentially many gives
\ESOnewbd{\mathrm{exp}}, and leaving them unbounded gives \ESOnew{}.

\subsection{Descriptive complexity}
\label{sec:background:dc}

From the descriptive point of view, a \emph{decision problem}~$P$ over a
vocabulary $\Ltt$ is an isomorphism-closed set of $\Ltt$-structures. It is
\emph{definable} in a logic~\Lmc{} if some \Lmc-sentence $\varphi$
over~$\Ltt$ has $\Att \mdl \varphi$ iff $\Att \in P$, for every finite
$\Ltt$-structure~$\Att$.
It is \emph{definable over ordered structures}, or definable in~$\Lmc(\leq)$,
if some \Lmc-sentence $\varphi$ over $\Ltt \cup \{\leq\}$ has $(\Att, \leq)
\mdl \varphi$ iff $\Att \in P$, for every finite $\Ltt$-structure~$\Att$ and
\emph{every} linear order~$\leq$ on its universe. The order is a device of the
definition and not part of the input: the problem is a set of
$\Ltt$-structures either way, and a sentence using~$\leq$ counts only if its
answer is the same whichever order it is given.

Let~\Cmc be a class of decision problems. A logic \Lmc{}
\emph{captures}~\Cmc if, for every vocabulary~$\Ltt$, the $\Ltt$-problems
in~\Cmc
are precisely the $\Ltt$-problems definable in~\Lmc{}, and it captures~\Cmc
\emph{over ordered structures} if they are precisely those definable
in~\Lmc{} over ordered structures. The characterizations this paper relies on
are collected in Table~\ref{tab:classes}, which lists them beside the
classes as the library defines them.

We use a construction on classes called \emph{exponential expansion}.
Similarly to a first-order interpretation~(\S\ref{sec:background:red}), it
defines a new structure on top of an old one, but here the new elements are
given by \emph{relations} (or tuples of relations) over the original universe,
cut down by a definable condition. Second-order sentences over the old
structure thus become first-order sentences over the new. As a universe of
size~$n$ carries $2^{n^a}$ relations of arity~$a$, the expanded universe is
exponentially larger and a resource bound read there is one exponential
higher. For a class~\Cmc, we write $\Cmc^{\mathrm{exp}}$ for the problems that
some exponential expansion turns into a problem of~\Cmc.

\begin{table}
  \caption{The complexity classes of the library: each is \emph{defined} by the
    (first) logic in the middle column, any further one being proved
    equivalent to it, and a \(\leq\) marking a fragment taken over ordered
    structures. The right-hand column gives a single problem proved complete
    for the class, not all. \PH{} has no complete problem unless the hierarchy
    collapses, and \GI{} is a degree~(\S\ref{sec:classes:degrees}), not
    captured by a logic.}
  \label{tab:classes}
  \small
  \setlength{\tabcolsep}{4pt}
  \begin{tabular}{lll}
    \toprule
    \textbf{Class} & \textbf{Definition} & \textbf{Complete problem} \\
    \midrule
    \Logspace   & $\FO({\leq}, \DTC)$      & \prob{det-reach} \\
    \NL         & $\SO\text{-Krom}({\leq})$, $\FO({\leq}, \TC)$  & \prob{reach} \\
    \Ptime      & $\SO\text{-Horn}({\leq})$, $\FO({\leq}, \LFP)$ & \prob{cvp} \\
    \NP         & \ESO, \ESOnewbd{d}       & \prob{sat} \\
    \coNP       & \ASO                     & \prob{taut} \\
    \DP         & \(\Sigma^1_1 \land \Pi^1_1\) & \prob{sat-unsat} \\
    \SigmaP{k}  & \(\Sigma^1_k\)           & \prob{qbf}\(_k\) \\
    \PiP{k}     & \(\Pi^1_k\)              & \prob{qbf}\(^\forall_k\) \\
    \PH         & \SO                      & -- \\
    \PSPACE     & \SO(\TC), $\FO({\leq}, \PFP)$, \expof{NL} & \prob{qsat} \\
    \EXPTIME    & \SO(\LFP), \expof{PTIME} & \prob{atm-space} \\
    \NEXPTIME   & \expof{NP}, \ESOnewbd{\mathrm{exp}} & \prob{wide-tiling} \\
    \EXPSPACE   & \SO(\PFP), \expof{PSPACE} & \prob{wide-corridor} \\
    \RE         & \ESOnew                  & \prob{finsat} \\
    \midrule
    \GI         & degree of \prob{graph-iso} & \prob{graph-iso} \\
    \bottomrule
  \end{tabular}
\end{table}

\subsection{First-order reductions}
\label{sec:background:red}

Reductions are defined through first-order interpretations, in the manner
of~\citet{dahlhaus1983reduction}.
Fix two vocabularies $\Ltt$ and $\Ltt'$. A $d$-dimensional first-order
interpretation $I$ from $\Ltt$ to $\Ltt'$ consists of a finite nonempty
set~$T$ of \emph{tags}, of a first-order $\Ltt$-formula~$\delta_t(\xol)$
(\emph{domain formula}) for
each tag~$t \in T$, where $|\xol|=d$, and, for each $k$-ary relation symbol
$R\in \Ltt'$ and each tuple $\tol = (t_1,\ldots,t_k) \in T^k$ of tags, of a
first-order $\Ltt$-formula~$\rho_{R,\tol}(\xol_1,\ldots,\xol_k)$, where
$|\xol_i| = d$. Applied to an $\Ltt$-structure~$\Att$, the interpretation
produces the $\Ltt'$-structure $I(\Att)$ whose universe holds one disjoint copy
of the admitted $d$-tuples per tag,
$\{(t,\aol) \mid t \in T,\ \aol\in\Delta_\Att^d,\ \Att\models\delta_t(\aol)\}$,
and whose relations are defined by $R^{I(\Att)}
  = \{((t_1,\aol_1),\ldots,(t_k,\aol_k))
      \mid \Att\models\rho_{R,\tol}(\aol_1,\ldots,\aol_k)\}$.
Textbook presentations have no tags and obtain the same copies from an order on
the input, raising the dimension and letting the extra coordinates take
distinguished values that the order names. We keep the tags explicit, as this
does not require the $\Ltt$-structure to be ordered.

\begin{figure*}
  \leanfile{snippets/problem.lean}
  \medskip
  \leanfile{snippets/reduction.lean}
  \caption{Decision problems, an isomorphism-closed property of relational
    structures, and first-order reductions between them, whose \decl{correct}
    field ranges over finite nonempty structures.}
  \label{fig:problem}
  \label{fig:reduction}
  \Description{Two Lean structure declarations. DecisionProblem has two
    fields: Holds, giving a proposition for each structure on a type, and
    iso\_invariant, requiring isomorphic structures to agree on it.
    FOReduction has a finite nonempty type of tags, a dimension, an
    interpretation built from the two, and a correctness field stating that
    the source problem holds of a finite nonempty structure exactly when the
    target problem holds of its image.}
\end{figure*}

A \emph{first-order reduction} from an $\Ltt$-problem $P$ to an $\Ltt'$-problem $Q$
is a first-order interpretation $I$ such that $\Att \in P$ iff $I(\Att) \in Q$
for every finite nonempty $\Ltt$-structure~$\Att$
(\S\ref{sec:framework:reductions} says why nonempty); we write $P \fored Q$,
which the Lean sources write \lean|P ≤ᶠᵒ Q|. Here the domain formulas are
trivial, $\delta_t \equiv \top$, so that the output universe is all of
$T \times \Delta_\Att^d$; only the relativized variant below uses them.
Our notions of hardness and completeness are the usual ones for this type of
reduction.
From the classical standpoint, such a reduction is computable by a uniform
family of constant-depth, polynomial-size Boolean circuits: $\FO({\leq})$
sits inside \ACz, which is $\FO({\leq}, \mathrm{BIT})$. The
containment is strict, and unconditionally so, a lower bound the library
proves (\S\ref{sec:classes:lower}).
A first-order reduction is therefore computable in polynomial time, and is in
particular a Karp reduction (proved in the library as
\decl{FOReduction.toLFP}, using the $\FO({\leq}, \LFP)$ characterization
of \Ptime{}).

Two relaxations of \fored{} are needed.
An \emph{order-invariant} reduction $P \foredord Q$ may use the symbol $\leq$
in its formulas and, order-invariantly as above, must be correct for every
linear order on the input. Any classical reduction that indexes its gadgets
by position is of this kind (\S\ref{sec:classes:catalog}).
A \emph{relativized} reduction $P \foredrel Q$ is an order-invariant one with
nontrivial domain formulas: its output universe is the definable set
$\{(t,\aol) \mid \Att \mdl \delta_t(\aol)\}$. Karp's
cover-testing gadget for Hamilton Circuit is of this kind.
The three relations are ordered by strength, \fored{} being the strongest,
and hardness travels forward along each of them, so a completeness proof may
use whichever is convenient. Which one a reduction needs is a property of its
gadget (see \S\ref{sec:problems:np}).

\subsection{First-order logic in Mathlib}
\label{sec:background:modeltheory}

Mathlib's \decl{ModelTheory} library formalizes classical first-order model
theory: syntax and semantics, substructures and elementary maps,
ultraproducts and compactness, Fra{\"i}ss{\'e} limits, types, and Skolem
functions. Its focus is not finite model theory: it has neither
Ehrenfeucht--Fra{\"i}ss{\'e} games nor logical characterizations of complexity
classes, and those are developed here. What the library builds on is its
first-order infrastructure: a vocabulary is a \decl{FirstOrder.Language}, an
\Ltt-structure on a type \Att an instance of \decl{L.Structure A}, a
formula a \decl{BoundedFormula} or a \decl{Sentence}, with Mathlib's
realization semantics, its notion of isomorphism, written \lean|A ≃[L] B|,
and its interface for reading a relation of a structure as its order~$\leq$.

Mathlib also has \decl{Set.Definable}, but it addresses a different question:
whether a subset of one fixed structure is defined by some formula, possibly
with parameters. Here a decision problem is an isomorphism-invariant property
of finite structures, and its defining formula must be retained. This is the
object required by the capture theorems and by the logical fragments developed
in the library.

\begin{figure*}
  \leanfile{snippets/interpretation.lean}
  \caption{First-order interpretation semantics: \decl{relFormula} determines
    \decl{Map}, whose interpreted universe is
    $\decl{Tag} \times (\decl{Fin}\,\decl{dim} \to \Att)$.}
  \label{fig:interpretation}
  \Description{Three Lean declarations: the FOInterpretation structure, whose
    one field sends each relation symbol and tuple of tags to a formula; the
    Map definition, whose interpreted universe is pairs of a tag and a
    dim-tuple of elements; and the mapStructure instance, which interprets a
    relation by realizing that formula.}
\end{figure*}

\section{The Reduction Framework}
\label{sec:framework}

We begin the tour of the library with the formalization of decision problems
and reductions defined in \S\ref{sec:background:dc}
and~\S\ref{sec:background:red}. Furthermore, we provide
a reliable way for the user to encode decision problems given in conventional,
algorithm-based terms into the language of descriptive complexity and to
decode the resulting statements back.

\subsection{Decision problems as predicates on structures}
\label{sec:framework:problems}

A vocabulary is a \decl{FirstOrder.Language}~\Ltt, which in Mathlib may carry
function and constant symbols as well, together with an \decl{IsRelational}
instance restoring the convention of~\S\ref{sec:background:fsl}.
Given such a vocabulary~\Ltt, a \decl{DecisionProblem L} is an
isomorphism-invariant predicate on \Ltt-structures. It is represented by a
structure with two fields: a predicate \decl{Holds} and a proof
\decl{iso_invariant} that isomorphic structures agree on it
(Figure~\ref{fig:problem}).

At this point the predicate \decl{Holds} is not restricted to
finite structures. Finiteness is imposed later by the notions that use it.
Thus, a user can formalize a
decision problem as its natural semantic predicate, without carrying
finiteness conditions through every definition. What a problem does on
infinite structures is then invisible to the theory:
membership and hardness are fields of a complexity class that come with
\decl{mem_congr_finite} and \decl{hard_congr_finite}, saying that two
problems agreeing on all finite structures agree on both.

Isomorphism-invariance says simply that renaming the elements of a structure
does not change whether it is a yes-instance. It is needed when reductions are
composed: the two-step output and the output of the composite have different
Lean carrier types, but are isomorphic. The invariant proof transfers the
answer between them.

\subsection{Interpretations}
\label{sec:framework:interpretations}

As in theory, a first-order reduction of an \Ltt-problem to an
$\Ltt'$-problem is a first-order interpretation from~\Ltt to~$\Ltt'$ that
respects the \Holds predicate. An interpretation is parameterized by a finite
type \decl{Tag} and by its dimension~$d$.
Its output elements are the tagged $d$-tuples, $\decl{Tag} \times
\Delta_\Att^d$.

These parameters fixed, an \decl{FOInterpretation} defines a single
field, \decl{relFormula} (Figure~\ref{fig:interpretation}). Fix a relation
symbol~\(R\) of~$\Ltt'$, of arity~\(n\), and
an assignment~$\tau$ of tags to
its \(n\) arguments. At that pair,
\decl{relFormula} supplies an \Ltt-formula~$\rho_{R,\tau}$ whose free
variables are the pairs~\((i,j)\) with \(1 \leq i \leq n\) and
\(1 \leq j \leq d\): the figure writes them
$\decl{Fin}\,n \times \decl{Fin}\,\decl{dim}$, counting from~\(0\). Writing the
arguments as in \S\ref{sec:background:red}, the variable~\((i,j)\) is the
\(j\)-th coordinate of~\(\xol_i\).
Applied to an \Ltt-structure \Att, these formulas define the
relations of the interpreted $\Ltt'$-structure \mbox{\lean|I.Map A|:} tagged
tuples \(\xol_1, \dots, \xol_n\) are related by~\(R\) exactly when
$\rho_{R,\tau}$ holds of their coordinates, $\tau$ being the tags they carry.
The library proves that interpretations map finite structures to finite
structures, nonempty structures to nonempty structures when their tag type is
nonempty, and isomorphic structures to isomorphic structures.

For a complete example, let $\Ltt = \Ltt' = \{E\}$ with~$E$ binary, and take
\(\decl{Tag} = \{\ell, r\}\) and \(d = 1\). An output element is then a
pair~\((t,u)\) of a tag and a vertex of~\Att, so \lean|I.Map A| carries a left
and a right copy of every vertex of~\Att. Since~$E$ is binary, an
assignment~$\tau$ is a pair~\((t_1,t_2)\) of tags, one per argument, and
\decl{relFormula} must be supplied at each of the four. As \(d = 1\), each
formula has one variable per argument, \((1,1)\) and~\((2,1)\), which we
write~\(x\) and~\(y\); the symbol \(R = E\) being fixed, we write
\(\rho_{t_1t_2}\) for the formula chosen at~\((t_1,t_2)\), so that
\(
  \rho_{\ell r} = E(x,y) \) and
\(
  \rho_{\ell\ell} = \rho_{r\ell} = \rho_{rr} = \bot
\).
The interpreted graph has an edge from the left copy of~\(u\) to the right
copy of~\(v\) exactly when \(E^\Att(u,v)\), and no other edges.
Appendix~\ref{sec:appendix:example} shows an interpretation of this shape
written out in Lean.

The definition of \S\ref{sec:background:red} differs in one more technical
respect. There, the domain formulas keep only the tagged tuples~\((t,\aol)\)
with \(\Att \mdl \delta_t(\aol)\). An \decl{FOInterpretation} has no such
field: its output universe is the whole tagged product
$\decl{Tag} \times \Delta_\Att^d$. Tagged tuples outside the intended universe
are harmless: the relation formulas can leave them isolated. This avoids
building a subtype of definable tuples into every interpreted structure. The
\(\delta_t\) come back as a field \decl{domFormula} in
\decl{RelFOInterpretation}, which is what the relativized reductions
$\foredrel$ are built from.

\subsection{Reductions}
\label{sec:framework:reductions}

\begin{figure*}
  \leanfile{snippets/encoding.lean}
  \caption{An \decl{Encoding} enforces no padding or compression}
  \label{fig:encoding}
  \Description{A Lean structure carrying, for each input, a size and a
    universe, with further fields elided, and two polynomial bounds: the
    universe is polynomial in the size, and the size polynomial in the
    universe.}
\end{figure*}
An \decl{FOReduction} from \Ptt to \Qtt, written $P \fored Q$, packages an
interpretation with a proof $\Ptt(\Att) \leftrightarrow \Qtt(I.\mathrm{Map}(\Att))$
for every finite nonempty input structure $\Att$
(Figure~\ref{fig:reduction}). In other words, it preserves the answer on
exactly the finite nonempty instances to which the library's complexity
classes attach membership and hardness statements.
We highlight that correctness is only required on nonempty inputs. A fixed-dimensional
interpretation maps an empty universe to an empty universe, and therefore cannot
produce a nonempty target instance from an empty source instance.

To explain further constructions we need one more technical definition. An
$\Ltt'$-formula~\(\varphi\) is \emph{pulled back} along an
interpretation~\(I\) from~\Ltt to~$\Ltt'$ into an \Ltt-formula~\(\varphi^I\):
every relation atom becomes the formula that~\(I\) supplies for that relation,
every quantifier over output elements becomes a block of~\(d\) quantifiers
over coordinates, one copy per tag, the tags being finite in number. The
pullback preserves meaning: \(\varphi^I\) is true in~\Att exactly
when~\(\varphi\) is true in \lean|I.Map A|
(\decl{FOInterpretation.realize_pull}).

Reductions can be composed, as usual. Given \(P \fored Q\) and \(Q \fored R\),
the library constructs \(P \fored R\) by pulling the formulas of the second
interpretation back along the first. The only Lean-specific issue is that the
carrier of the composite is not definitionally the same type as the carrier
obtained by applying the two interpretations in sequence: one is a flattened
tuple of coordinates, the other a tuple of tuples. The library constructs an
isomorphism between them, then invokes the target problem's
\decl{iso_invariant} proof. Hence \decl{FOReduction.trans} establishes
transitivity. Together with the
one-dimensional, one-tag identity interpretation, this makes $\fored$ a
preorder.

Definability travels backward along the same pullback: if~\(\varphi\)
defines~\(Q\) and \(I\) witnesses \(P \fored Q\), then \(\varphi^I\)
defines~\(P\). Its contrapositive (\decl{not_le_of_not_foDefinableFree}) is
the lower-bound form: one inexpressibility result rules out every reduction to
an \FO-definable problem.

The two relaxations of \S\ref{sec:background:red} are defined analogously.
An \decl{OrderedFOReduction}, written \lean|P ≤ᶠᵒ[≤] Q|, interprets over the
vocabulary expanded with~$\leq$ and must be correct for every linear order on
the input; a \decl{RelOrderedFOReduction}, \lean|P ≤ʳᶠᵒ[≤] Q|, is built on a
\decl{RelFOInterpretation} and so carries the domain formulas as well. Both
compose, and definability travels backward along them too.

\subsection{Encodings and decodings}
\label{sec:framework:encodings}

Decision problems in the library concern finite structures, whereas users
usually start from concrete data: lists, formulas, weights, and the like. An
\emph{encoding} translates such data into structures, allowing the library's
definability and complexity results to apply; a \emph{decoding} reads
results back.

\decl{Encoding} records the input type and its conventional size measure, the
encoded universe and relations as computations, and two polynomial bounds
relating the cardinality of that universe to that size. Both directions are
needed: representing binary integers over a unary universe makes the structure
exponential in the input length, and compression the other way can invalidate
a hardness result. Both are fields (Figure~\ref{fig:encoding}).
Preservation of meaning is a separate predicate,
\decl{Encoding.Faithful}, saying that a concrete instance
and its encoded structure are decided alike; keeping it apart lets one
encoding serve several problems over the same vocabulary.

A faithful encoding transfers membership: if the structural problem~\Ptt lies
in a class, so does the user's. Hardness does not transfer in the same way, as
\Ptt may be hard because of malformed structures that no instance encodes. The
library therefore restricts \Ptt to a well-formedness condition~\Wtt,
obtaining $\Wtt \cap \Ptt$, and asks for a \decl{Decoding}: a computation
taking a finite-structure presentation to the concrete instance it stands for,
or to \lean|none| when it stands for none, agreeing with~\Ptt where it
succeeds and defined on every well-formed nonempty presentation. Hardness of
the restricted problem then concerns only structures that some instance
presents.

\begin{figure*}
  \leanfile{snippets/class.lean}
  \caption{Constructor for \decl{ComplexityClass}:
    a membership predicate, and the three obligations it must meet}
  \label{fig:class}
  \Description{The signature of a Lean definition that takes a membership
    predicate on decision problems together with three proofs about it, namely
    that membership travels backward along plain first-order reductions,
    backward along ordered ones, and depends only on finite structures, and
    returns a complexity class.}
\end{figure*}
\section{Complexity Classes, Logically Defined}
\label{sec:classes}

A class here is a set of decision problems closed under first-order reductions
and usually obtained from a logic.

\subsection{Defining a complexity class}
\label{sec:classes:closure}

A class is defined by the constructor \decl{ComplexityClass.ofMem}, from a
membership predicate together with two proofs about it: that membership is
stable under first-order reductions, and that it is determined by the finite
structures (Figure~\ref{fig:class}).
The stability proof comes from a \emph{pullback lemma}: if the target problem
is definable in the logic, so is the problem reducing to it. Being an argument
of the constructor, it must be in hand before the class exists, and there is
one such lemma per logic (Appendix~\ref{sec:appendix:logics}), because the
pullback of \S\ref{sec:framework:reductions} is first-order work only for
\FO{}. A
relation variable of arity~$n$ in the target sentence ranges over relations
on the output universe $\decl{Tag} \times \Delta_\Att^d$, which no relation
on~$\Delta_\Att$ is. It is replaced by a \emph{block} of $|\decl{Tag}|^n$
relation variables of arity~$nd$, one per assignment of tags to its
arguments, and every atom of the sentence is rewritten to read the variable
its tags select (\decl{SOBlock.pull}). The rewriting is atom by atom, so it
takes a clause to a clause and preserves the Horn and Krom fragments (\decl{SigmaSOHornDefinable.of_orderedReduction},
\decl{SigmaSOKromDefinable.of_orderedReduction}): this is why \Ptime{} and
\NL{} are closed without leaving their defining fragments. A fixed-point
operator is pulled the same way, its relation variable as a block and its
step formula as first-order (\decl{LFPDefinable.of_orderedReduction}, and
likewise for \IFP{}, \PFP{} and~\TC{}).
Hardness is then determined by the membership predicate:
$P$ is hard for the class when every member of it reduces to~$P$
(\decl{cofinalHard_iff}).
Completeness for a class is a pair of certificates, for membership
and for hardness.
Each class of Table~\ref{tab:classes} has a problem proved complete for it
(\S\ref{sec:problems}); \PH{} has none unless the polynomial hierarchy collapses, hence
the dash in the table.

\begin{toappendix}
\section{How Each Logic Is Represented in Lean}
\label{sec:appendix:logics}

Mathlib supplies first-order syntax and semantics; it has no second-order
quantifier and no fixed-point or closure operator, and the library adds none
as syntax. Each logic is a \emph{definability
predicate} on \decl{DecisionProblem}: a type of data, a sentence, a program
or a specification; a Lean definition giving each datum a meaning on each
structure; and the predicate itself, that some datum agrees with the problem
on every finite nonempty structure. Two conventions run through all of them.
A logic taken over ordered structures reads the vocabulary
\lean|L.sum Language.order| under a \lean|[LinearOrder A]| instance, and the
agreement is required for \emph{every} linear order, which is what makes the
notion order-invariant and the predicate a property of the problem alone;
the order-free variants (\decl{FODefinableFree}, \decl{IFPDefinableFree},
\decl{PFPDefinableFree}) drop the expansion and the instance. And each
predicate comes with a lemma \lean|_congr|, saying it depends only on the
finite structures, and a lemma \lean|.of_orderedReduction|, pulling it back
along an ordered reduction: the obligations of \decl{ComplexityClass.ofMem}
(Figure~\ref{fig:class}), the plain reduction being an ordered one that
ignores the order.

\paragraph{First-order logic.} \decl{FODefinableFree} asks for a Mathlib
\lean|Sentence| over the vocabulary itself, \decl{FODefinable} for one over
its ordered expansion, and \decl{AC0Definable} for one over the arithmetic
expansion, where $\leq$, $+$ and~$\times$ are relation symbols computed from
the order.

\paragraph{Second-order logic.} A quantifier block is a finite index type of
relation variables with their arities; its vocabulary \lean|B.lang| has one
relation symbol per variable, and an assignment gives each variable a
relation on the universe. A list of blocks is folded into the base vocabulary
by \lean|soLang|, and \lean|SORealize| quantifies the blocks alternately,
existentially first or universally first, over a first-order kernel in the
summed vocabulary. So a $\Sigma^1_k$ sentence is a list of $k$ blocks and a
Mathlib sentence, and \decl{SigmaSODefinable}~$k$ and \decl{PiSODefinable}~$k$
are the two polarities. Pulling a block back along an interpretation
(\S\ref{sec:classes:closure}) is a construction on blocks alone,
\decl{SOBlock.pull}: an index becomes a pair of an index and a tuple of tags,
an arity is multiplied by the dimension.

\leanfile{snippets/logic-block.lean}

The Horn and Krom fragments are not recognized on sentences but built as
data. A \decl{HornClause} is a first-order guard over the input vocabulary, a
list of positive second-order body atoms, and an optional head; a program is
a list of clauses, universally quantified over $k$ shared first-order
variables. \decl{SigmaSOHornDefinable} asks for a block and a program, over
the ordered vocabulary, such that the problem holds exactly when some
assignment satisfies every clause; \decl{SigmaSOKromDefinable} is the same
with clauses of at most two signed second-order literals. Value invention
(\decl{SigmaSONewDefinable}) extends the vocabulary by the unary mark
\lean|old|, and reads the block over \lean|A ⊕ Fin m|, the invented
elements related to nothing, for some \lean|m : ℕ| existentially quantified
in Lean and not in the logic, which is what leaves the number of invented
elements unbounded.

\paragraph{Fixed points.} An \decl{LFPDef} is a block, a list of Horn rules
defining its variables, and an \emph{unrestricted} output sentence over the
ordered vocabulary expanded by the block; it holds when the output is true at the
least fixed point of the rules, which is the inductive predicate
\decl{Derives} of derivable atoms. Negating the output negates the value,
so closure under complement (\decl{LFPDefinable.compl}) is one line, and
Gr{\"a}del's theorem (\S\ref{sec:classes:relations}) is the translation
between an \decl{LFPDef} and a Horn program, in both directions.

\leanfile{snippets/logic-lfp.lean}

The inflationary and partial fixed points share their data: a
\decl{StepDef} is a block, one first-order step formula per variable, over
the vocabulary expanded by the block and with the variable's arguments as
free variables, and an output sentence. The step formulas are unrestricted,
and the two logics differ only in the iteration they apply: \decl{IFPHolds}
accumulates each stage into the previous one and reads the output at the
limit; \decl{PFPHolds} iterates the bare step from the empty assignment and
holds when some stage is a fixed point at which the output is true, so a
diverging iteration makes the definition false whatever its output. The
textbook reading of \S\ref{sec:background:fsl}, which returns the empty
relations instead, differs only when the output holds at the empty
assignment; \decl{StepDef.realize_pfpValue_iff} compares the two, and a
definition is carried from one reading to the other by guarding its output
with ``the stage is a fixed point'', a first-order condition. The
convention is chosen because it makes the translation to $\SO(\TC)$ direct:
acceptance of the walk \emph{is} ``some stable stage satisfying the output
is reachable'', with no divergence detection, which unordered structures
could not supply.

\leanfile{snippets/logic-step.lean}

\paragraph{Transitive closure.} A \decl{TCSpec} is a walk on $k$-tuples
carrying a finite \emph{mode}: a transition formula per pair of modes, with
the current tuple and the next as free variables, and start and accepting
formulas per mode, all over the ordered vocabulary. A node is a mode with a
tuple, the transition formulas define the edges, and the specification
accepts when some accepting node is reachable from some starting node;
\decl{TCDefinable} asks the problem to coincide with acceptance. The mode is
to a walk what a tag is to an interpretation: a one-element universe has one
tuple, so finite state cannot live in the tuple. \decl{DTCDefinable} replaces
the transition by its determinization \decl{TCSpec.det}, ``this step, and
no other step out of the current node'', so that determinism is a formula
and not a side condition on the specification. $\SO(\TC)$ is the same walk on
assignments of a block (\decl{SOTCSpec}), its transition a sentence over two
copies of the block.

\leanfile{snippets/logic-tc.lean}

\paragraph{Exponential expansions.} An \decl{ExpExpansion} is the
exponential expansion of \S\ref{sec:background:dc} as a Lean
\lean|structure|: a finite type of tags, a block whose assignments are the
points, a domain sentence~$\delta_t$ per tag, and, for each relation symbol
of the expanded vocabulary and each tuple of tags, a defining sentence over
as many copies of the block as the symbol has arguments. Its \lean|Map|
sends~\Att{} to the structure whose universe is
$\{(t, \rho) \mid (\Att, \rho) \mdl \delta_t\}$, and
\decl{ExpDefinable}~$\Cmc$ holds of~$P$ when some expansion turns it into a
member of~$\Cmc$. \decl{ComplexityClass.exp} is that predicate handed to
\decl{ComplexityClass.ofMem}; \decl{SOLFPDefinable} and \decl{SOPFPDefinable},
the logics of \EXPTIME{} and \EXPSPACE{} in Table~\ref{tab:classes}, are
\decl{LFPDefinable} and \decl{PFPDefinable} read through an expansion.

\leanfile{snippets/logic-exp.lean}

\paragraph{Games.} The Ehrenfeucht--Fra{\"i}ss{\'e} game is the predicate
\decl{efStage}, by recursion on the number of rounds: a position is a pair
of tuples, one in each structure, and survives zero rounds when it is a
partial isomorphism, $n+1$ rounds when it is one and every element chosen on
either side can be answered on the other so that the extended position
survives~$n$. \decl{EFEquiv} plays from the empty position, and the one
lemma the lower bounds rest on, \decl{realize_efStage}, says that a formula
of quantifier rank at most~$n$ cannot separate the two sides of a position
surviving $n$ rounds. The strategies are then combinatorics: on bare sets
of at least $n$ elements (\decl{efEquiv_bare}) and on linear orders of at
least $2^n$ elements (\decl{efEquiv_linearOrder}), whence
\decl{even_not_foDefinableFree} and \decl{even_not_foDefinable}.

\leanfile{snippets/logic-games.lean}

The $k$-pebble game is kept apart from the logic. \decl{EquivK} is the
greatest fixed point of one round of pebble exchange over an abstract
initial relation, instantiated at agreement on the atomic type over the
finitely many symbols a definition mentions; \decl{realize_equivK} is the
$k$-variable invariance lemma, its budget hypothesis standing in for a
syntactic $k$-variable fragment, which is never defined. Each stage of an
induction expands the structure by an invariant relation and leaves
\decl{EquivK} unchanged (\decl{equivK_inf_eq}), so an inflationary
definition whose formulas fit in $k$ variables cannot separate two
$\equiv^k$-equivalent structures (\decl{StepDef.ifpHolds_equivK₂}), and bare
sets of $2k+2$ and
$2k+3$ elements are such a pair, whence
\decl{exists_mem_PTIME_not_ifpDefinableFree}.

\end{toappendix}

\subsection{A catalog of classes}
\label{sec:classes:catalog}

Most definitions through logics are standard, and we
cite~\citet{immerman1999descriptive} by theorem number wherever the book has
the result. \Logspace{} is $\FO({\leq}, \DTC)$ (Thm.~9.11); \NL{} is
$\FO({\leq}, \TC)$ (Cor.~9.22) and $\SO\text{-Krom}({\leq})$ (Thm.~9.32);
\Ptime{} is $\FO({\leq}, \LFP)$ (Thm.~4.10),
after~\citet{immerman1986relational} and~\citet{vardi1982complexity}, and
$\SO\text{-Horn}({\leq})$ (Thm.~9.32), the two second-order fragments being due
to~\citet{gradel1992capturing}. \NP{} is \ESO{} by Fagin's
theorem~\cite{fagin1974generalized} (Thm.~7.8); the levels \SigmaP{k}
and~\PiP{k} are $\Sigma^1_k$ and~$\Pi^1_k$ (Thm.~7.21),
after~\citet{stockmeyer1976polynomial}, and \PH{} is \SO{} (Cor.~7.22).
\PSPACE{} is $\FO({\leq}, \PFP)$ (Thm.~10.13), due
to~\citet{vardi1982complexity}, and $\SO(\TC)$ (Cor.~10.29). \DP{} is
from~\citet{papadimitriou1984complexity}, read as
$\Sigma^1_1 \wedge \Pi^1_1$ from Fagin's theorem.

The last four rows are of a different kind, and the book has none of them.
Several are reached by the exponentiation operator $\Cmc \mapsto \expof{\Cmc}$
(\S\ref{sec:background:fsl}), which the library proves to relate the classes
one exponent apart: $\PSPACE = \expof{NL}$ (\decl{PSPACE_eq_NL_exp}),
$\EXPTIME = \expof{PTIME}$, $\EXPSPACE = \expof{PSPACE}$. \EXPTIME{} and
\EXPSPACE{} are nevertheless \emph{defined} by $\SO(\LFP)$ and $\SO(\PFP)$,
after the fixpoint logics of~\citet{abiteboul1997fixpoint}. \NEXPTIME{} is
third-order in the literature~\cite{leivant1989descriptive}, a syntax the
library lacks, so it is $\expof{NP}$ here, with \ESOnewbd{\mathrm{exp}} for
its logic. \RE{} is \ESOnew{}, unbounded value invention, the device of the
object-creating query languages of~\citet[ch.~18]{abiteboul1995foundations},
its complete problem \prob{finsat} being
Trakhtenbrot's~\cite{trakhtenbrot1950}.

\label{sec:classes:degrees}
Not every class need come from a logic. The \emph{degree}
\decl{ComplexityClass.below} of a problem~$P$, the problems that reduce
to~$P$, is itself a class, closed under reductions by construction, so
completeness becomes statable with no logic anywhere. Any class with a
complete problem is the degree of that problem, and where both apply the two
agree, as in \decl{NP_eq_below_sat} and \decl{PTIME_eq_below_hornSat}:
\prob{sat}-hardness \emph{is} \NP-hardness as a theorem.


\subsection{Relations between the classes}
\label{sec:classes:relations}

The library proves several structural results about the classes of
Table~\ref{tab:classes}. Many inclusions follow directly from the logical
definitions. For example,
\NL{} contains \Logspace{} because a deterministic transitive closure is a
transitive closure (\decl{LOGSPACE_subset_NL}); \NP{} and \coNP{} sit inside
\DP{} because an \ESO{} or \ASO{} condition is a conjunction of
both with a tautology (\decl{NP_subset_DP}, \decl{coNP_subset_DP}); \DP{} sits
inside each level of the polynomial hierarchy by the same argument
(\decl{DP_subset_sigmaP_two}, \decl{DP_subset_piP_two}). For $\NL \sbs \Ptime$
we also use reductions and hardness: \prob{2-sat}, a complete problem for
\NL{}, is defined by a Horn second-order sentence and thus belongs to
\Ptime{}, implying the inclusion of the whole class as the degree of
\prob{2-sat}. The polynomial hierarchy is handled uniformly in $k$: four step
lemmas $\SigmaP{k} \subseteq \SigmaP{k+1}$, $\SigmaP{k} \subseteq \PiP{k+1}$,
$\PiP{k} \subseteq \SigmaP{k+1}$, $\PiP{k} \subseteq \PiP{k+1}$, each level
included into~\PH. Likewise \decl{QBF_complete} gives completeness for every
level in one statement, after~\mbox{\citet{wrathall1976complete}.}

\paragraph{Capture theorems.} We prove Gr{\"a}del's
equivalence~\cite{gradel1992capturing}: the Horn
fragment of \ESO{} coincides with $\FO(\LFP)$ on ordered structures,
because the Horn rules can be evaluated by iterating a least fixed point, and
conversely any LFP definition can be written as a Horn program by making the
iteration explicit in the rules (\decl{lfpDefinable_iff_sigmaSOHornDefinable}).
Read against the definition of \Ptime{} as $\SO\text{-Horn}({\leq})$, this is
the Immerman--Vardi
theorem~\cite{immerman1986relational,vardi1982complexity}, that $\FO(\IFP)$
over ordered
structures captures \Ptime{} (\decl{lfpDefinable_iff_mem_PTIME}). The latter
uses the equivalence $\FO(\IFP) = \FO(\LFP)$ for ordered structures, where the
least and inflationary fixed points coincide: the order lets us detect
convergence (\decl{ifpDefinable_iff_lfpDefinable}). The partial fixed point
captures \PSPACE{}: $\FO(\PFP)$ over ordered structures equals \PSPACE{},
proved by showing $\FO(\PFP) = \SO(\TC)$
(\decl{pfpDefinable_iff_sotcDefinable}). These theorems connect the
fragment-based definitions (\SO-Horn for \Ptime{}, $\SO(\TC)$ for \PSPACE{})
with the operator-based ones ($\FO(\IFP)$ for \Ptime{}, $\FO(\PFP)$ for
\PSPACE{}): \decl{ifpDefinable_iff_mem_PTIME},
\decl{pfpDefinable_iff_mem_PSPACE}.

\paragraph{Closure under complementation.}
The class \NL{} is closed under complementation by the
Immerman--Szelepcs{\'e}nyi
theorem~\cite{immerman1988nondeterministic,szelepcsenyi1988method}. In the
library, it is proved for $\FO(\TC)$ via inductive counting
(\decl{TCDefinable.compl}), and the Krom fragment inherits this through the
translation to $\FO(\TC)$ (\decl{sigmaSOKromDefinable_compl_iff}); at the level
of classes this is \decl{NL_eq_coNL}.
Then, \Ptime{} is closed because $\FO(\LFP)$ is closed under negation
(\decl{LFPDefinable.compl}); and \PSPACE{} is closed because every $\SO(\TC)$
problem reduces to \prob{qsat}~\cite{stockmeyer1973word}, flipping the answer
of which gives the
complement within $\SO(\TC)$ (\decl{PSPACE_eq_coPSPACE}).
Further, \EXPTIME{} and \EXPSPACE{} inherit closure from \Ptime{} and \PSPACE{}
via the exponential operator $\Cmc \mapsto \expof{\Cmc}$
(\decl{ComplexityClass.exp}), which commutes with complement
(\decl{ComplexityClass.exp_compl}).

\paragraph{Abiteboul--Vianu theorem.} This result is of a different nature, an
equivalence with an open question on either side: $\FO(\IFP)$ and
$\FO(\PFP)$ coincide over \emph{unordered} finite structures exactly when
$\Ptime = \PSPACE$~\cite{abiteboul1995computing,abiteboul1991generic},
\decl{ifpDefinableFree_eq_pfpDefinableFree_iff_ptime_eq_pspace}. We follow the
purely logical route of~\citet{dawar1995infinitary}, whose invariant structure
and its definable order replace the relational machine of the original proof.

\subsection{Unconditional lower bounds}
\label{sec:classes:lower}

Ehrenfeucht--Fra{\"i}ss{\'e} games~\cite{ehrenfeucht1961application} are the
tool the machine-first developments have no analogue of: they argue about what
a \emph{logic} cannot say. The witness throughout is \prob{even}, whether a
finite set has an even number of elements. It is not first-order definable,
even order-invariantly (\decl{even_not_foDefinable}).

{\sloppy It is one walk along an order, though, so
\(\FO({\leq}) \subsetneq \FO({\leq}, \TC)\) outright
(\decl{exists_tcDefinable_not_foDefinable}); and arithmetic defines it too,
which separates $\FO({\leq})$ from \ACz{}, read here as the logic
$\FO({\leq}, +, \times)$ and not as a circuit class
(\decl{exists_ac0Definable_not_foDefinable}).\par}

The $k$-pebble game carries \prob{even} to the fixed-point
logics~\cite[ch.~7]{ebbinghaus1995finite}, implying that order-free \FO(\IFP)
does not capture \Ptime{} (\decl{exists_mem_PTIME_not_ifpDefinableFree}). The
\(\leq\) in every capture theorem is doing work.

\section{Machine Bridges}
\label{sec:machines}

\begin{figure*}
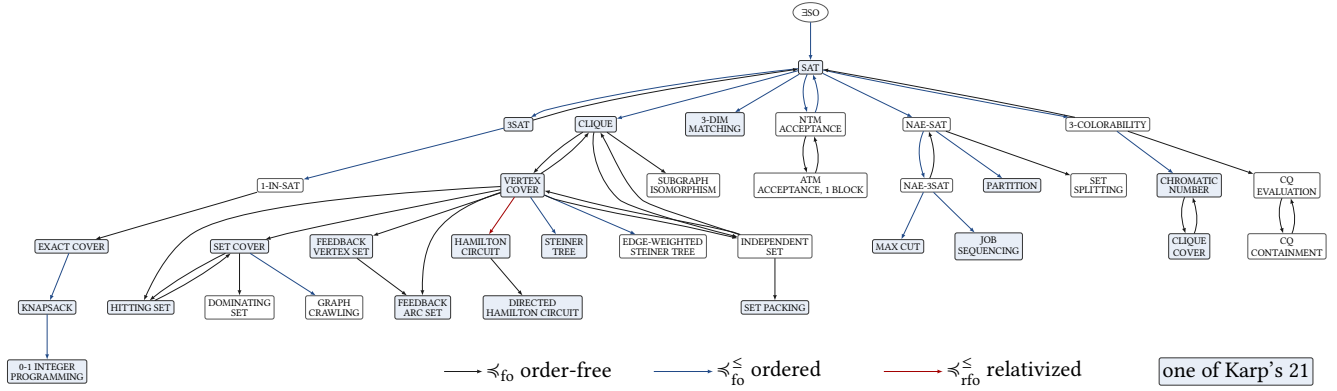

  \centering
  \resizebox{\textwidth}{!}{\dotfigure{figures/np-reductions}}
  \caption{Problems the library proves \NP-complete, and reductions
    between them.
    Karp's 21 are highlighted. An arrow between problems is a proved first-order reduction,
    colored by the form of the reduction.
    The graph is not strongly
    connected only because
    \emph{membership} in \NP is proved through \ESO{} definability rather
  than through a reduction.}
  \label{fig:reductions}
  \Description{A directed graph whose nodes are the problems the library
    proves NP-complete and whose edges are the first-order reductions proved
    between them. Shaded boxes mark Karp's twenty-one problems. Edge color
    gives the form of the reduction, black for order-free, blue for ordered
    and red for relativized, as the key along the bottom records. An
    existential second-order node feeds SAT, membership in NP being proved by
    definability rather than by a reduction.}
\end{figure*}
The \emph{machine bridges} link the logically defined classes to the standard
definitions by Turing machine. For each class we define, in the library's
vocabulary, the canonical \emph{acceptance problem}; for \NP{} an instance
describes a nondeterministic machine together with its input, and the question
is whether it accepts within as many steps as the instance has elements. We
then prove that \NP{}, defined by \ESO-definability, coincides with the degree
of that problem (\decl{mem_NP_iff_le_ntmAccept}). No machine model is declared
as a Lean type: the machine is \emph{data in the instance}.

\subsection{Acceptance as a decision problem}
\label{sec:machines:problem}

The vocabulary \lean|FirstOrder.Language.turing| encodes a computation as a
finite structure. Two unary symbols sort the universe, the \emph{positions} --
tape cells and time steps at once -- and the \emph{transitions}; the remaining
symbols mark a start state, the accepting states, the blank and the rightward
transitions, order the positions into a tape by \texttt{le}, give their initial
contents, and tie each transition to the states it runs between and the symbols
it reads and writes. States and tape symbols get no sort of their own: no part
of the definition quantifies over them.

\decl{NTMAccept} then asks whether the machine an instance describes accepts
within as many steps as there are positions. It holds when the structure is
well-formed -- the tape is ordered linearly, there is at least one position,
the input is functional, the blank unique -- and some run reaches an accepting
configuration in time. \decl{DTMAccept} adds a determinism guard; the
space-bounded
\decl{NTMAcceptSpace} and \decl{DTMAcceptSpace} drop the step bound, acceptance
becoming reachability in the configuration graph. Since positions serve as time
steps, the budget is a count of universe elements and the vocabulary needs no
arithmetic: a $d$-dimensional reduction (\S\ref{sec:framework:reductions})
builds $n^d$ of them, which is where the polynomial bound comes from.

\subsection{Recovering the machine classes}
\label{sec:machines:capture}

Hardness is the easy half: \prob{sat} is complete for \NP{}, so it suffices to
reduce it to \decl{NTMAccept}, building a machine that guesses an assignment
and checks clauses, and \Ptime{} enters the degree of \decl{DTMAccept} the same
way from \prob{horn-sat}. For membership, \decl{NTMAccept} is in \NP{} by
writing the run as an \ESO{} sentence: one existential block guesses the state,
head position and tape contents at each step, and a first-order kernel checks
that they form a valid accepting run. Determinism makes the run of
\decl{DTMAccept} unique, so it is derivable by a Horn program and the problem
lands in \Ptime{}. Each direction closes into a biconditional:
\decl{mem_NP_iff_le_ntmAccept}, \decl{mem_PTIME_iff_le_dtmAccept}.
This is how Fagin's theorem is proved. Since \NP{} is \emph{defined} as
\ESO{}, it is not an equality of classes; it is instead the fact that \ESO{}
is exactly the degree of nondeterministic polynomial-time acceptance.

Below \Ptime{} the bridge is a capture rather than a complete problem. A
logarithmic-space machine has polynomially many configurations, each a tuple
over the instance, so its configuration graph is definable on the instance and
acceptance \emph{is} reachability, leaving nothing to complete. \Logspace{} and
\NL{} are exactly what two-way multi-head automata recognize
(\decl{mem_LOGSPACE_iff_automaton}, \decl{mem_NL_iff_automaton}), and \ACz{}
what constant-alternation logarithmic time does
(\decl{ac0Definable_iff_ltDecidable}).

{\sloppy Above \Ptime{} the run outgrows the instance. The bridge there keeps
the complete problem of \NP{} and \Ptime{}.
The space-bounded
\decl{NTMAcceptSpace} and \decl{DTMAcceptSpace} are complete for \PSPACE{}
(\decl{spaceMachines_PSPACE_complete}), and \EXPTIME{} is reached by the
space-bounded alternating machines of~\citet{chandra1981alternation}
(\decl{atmAcceptSpace_EXPTIME_complete}). \NEXPTIME{} and \EXPSPACE{} need
\emph{wide machines}, whose universe is the power set of the instance together
with the instance itself, so that tape and clock run to~$2^n$
(\decl{wideRegAccept_NEXPTIME_complete},
\decl{wideAcceptSpace_EXPSPACE_complete}).
Appendix~\ref{sec:appendix:machinelevels} details each level.\par}

For \RE{} no machine appears on either side. An \ESOnew{} problem is
semi-decidable because its invented values can be found by unbounded search
(\decl{RE_subset_rePred}); conversely a semi-decidable problem reduces to the
\RE-complete halting problem. Together they give \decl{mem_RE_iff_rePred},
in Mathlib's own \decl{REPred} sense.

\subsection{Cook--Levin in its textbook form}
\label{sec:machines:cooklevin}

A reduction from \decl{NTMAccept} to \prob{sat} is the
classical Cook--Levin theorem~\cite{cook1971complexity,levin1973universal}:
encode the run as an \ESO{} sentence, then introduce a fresh variable per
subformula with clauses enforcing the equivalence (Tseitin
construction~\cite{tseitin1968complexity}). That is
\decl{ntmAccept_interreducible_sat}, both reductions in one statement.
\decl{SAT_complete_for_ntmAccept} is the same content with hardness quantified
over the class, i.e., \prob{sat} complete for the \NP{} the \emph{machine}
defines. This is what the developments compared in
\S\ref{sec:related:cooklevin} prove.

\begin{toappendix}
\section{What the Machine Characterization Costs at Each Level}
\label{sec:appendix:machinelevels}

Table~\ref{tab:machinelevels} applies the metric of
Table~\ref{tab:cooklevin} along the hierarchy instead of across provers: each
number is the declaration closure of a theorem, counted in the library's own
modules, with the rows and the order of Table~\ref{tab:classes}. The two rows
below \Ptime{} carry no membership and hardness split because the bridge there
is a capture rather than a complete problem (\S\ref{sec:machines:capture}).

The \emph{marginal} column is a difference of declaration \emph{sets}, not of
totals, and it is taken over the entry in the \emph{over} column, what the
library had already paid for when the row was added. For most rows that is the
class's problem in Table~\ref{tab:classes}, the machine being the second
complete problem for its class: \prob{cvp} for \Ptime{}, \prob{sat} for
\NP{}. At the three exponential classes the machine comes first instead, the
tilings being drawn out of it afterwards, so baselining on a tiling would
measure backwards; the baseline is the machine of the class each one is the
exponential of -- \decl{DTMAccept} for \EXPTIME{} -- and the marginal measures
what reading the same machine over an exponential expansion costs. A
\(\dagger\) row is taken over the first machine of its class.
The marginal can be far larger than the gap between the totals, because the
closures overlap only in part. The closure of \decl{ATMAcceptSpace} exceeds
that of \decl{DTMAccept} by \mlEXPTIMETotalOver{} lines, and yet
\mlEXPTIMEMarginal{} of its lines are ones the deterministic machine's proof
never touches.

The cost of machine characterization is often comparable to that at the
\NP{} level. In particular,
every row from \Ptime{} to \EXPTIME{}, and
\RE{} above them, closes below \gkOwnLines{}~lines, the smallest Cook--Levin
closure of any other library in Table~\ref{tab:cooklevin}.

The \(\dagger\) rows are cheap: a second machine for a
class is reached from the first and not from the logic.
\mlPSPACEnondetMarginal{}~lines buy the nondeterministic space machine over the
deterministic one, the deterministic side being proved hard and transferred, so
that no Savitch argument~\cite{savitch1970relationships} is needed on the
machine side; \mlEXPSPACEdetMarginal{}~lines buy the deterministic wide machine
over the nondeterministic one, and \mlNPaltMarginal{} the one-block alternating
machine over \decl{NTMAccept}.

The two \emph{wide} rows are the exception, at \mlNEXPTIMETotal{} and
\mlEXPSPACETotal{} lines against \mlEXPTIMETotal{} for \EXPTIME{}, which is
exponential too. The construction sets them apart: a reduction
into a wide machine quantifies over an exponential expansion of the instance
(\S\ref{sec:machines:capture}), so its interpretation has to write the
machine's program down inside the logic instead of naming it, and the hardness
half alone runs to \mlEXPSPACEHard{}~lines.

\begin{table*}
  \caption{Cost of the machine characterization, class by class, with the rows
    of Table~\ref{tab:classes}. Each number is a \emph{declaration closure} in
    the sense of Table~\ref{tab:cooklevin}: the declarations the theorem's
    proof term transitively reaches, counted as the source lines they occupy,
    so a module contributes only the part of itself the proof uses. Only the
    library's own modules are counted, not the Mathlib declarations beneath
    them. \emph{Membership} and \emph{Hardness} are the closures of the two
    halves and overlap both each other and \emph{Complete}, so the three do
    not add up. \emph{Marginal} is
    $\mathrm{closure}(\emph{Complete}) \setminus \mathrm{closure}(\emph{over})$
    on declaration sets, not a difference of totals; \emph{over} is what was
    already paid for, a prototypical complete problem of the class where there
    is one and a comparable machine result otherwise. A dash means the
    theorem has no separately named halves.
    \textsuperscript{$\ast$}\,A capture biconditional, not a complete
    problem, so it has no membership/hardness split.
    \textsuperscript{$\dagger$}\,A second machine for a class already rowed
    above.}
  \label{tab:machinelevels}
  \small
  \setlength{\tabcolsep}{5pt}
  \begin{tabular}{llrrrrrl}
    \toprule
    \textbf{Class} & \textbf{Machine statement} & \textbf{Mod.}
      & \textbf{Membership} & \textbf{Hardness} & \textbf{Complete}
      & \textbf{Marginal} & \textbf{over} \\
    \midrule
    \machinelevelrows
    \bottomrule
  \end{tabular}
\end{table*}

\end{toappendix}

\section{A Catalog of Complete Problems}
\label{sec:problems}

The library proves \dcCompleteThms{} completeness theorems across
\dcCompleteClasses{} classes, counted by class in
Appendix~\ref{sec:appendix:completeproblems}. Most are \NP{} or \coNP{}; the
rest follow a few patterns, and one problem fits no logically defined class.

Adding a problem costs little once the framework is there. Setting aside the
machine models, whose modules Appendix~\ref{sec:appendix:machinelevels} costs
instead, the median problem of the catalog is \problemMedian{} lines of its own
modules, with quartiles \problemQOne{} and~\problemQThree{}
(Appendix~\ref{sec:appendix:problemcosts} costs them one by one). Those are
source lines of what a problem adds, since we are measuring the cost of adding
another problem onto a base library.

Part of that cost is notation. Mathlib's \decl{BoundedFormula} is locally
nameless: a variable bound by the $k$-th enclosing quantifier block is
\lean|Sum.inr i| under $k-1$ applications of \lean|Sum.inl|, so writing a
clause means counting blocks, and reading one means counting them back. The
library's \lean|fo
throughout, names the variables and applies a symbol to them as
\lean|R(x, y)|. It is a macro, not a new type of formulas: it elaborates to
exactly the term one would have written, so a definition converted to it keeps
its equation lemma, and every \lean|simp only| proof about it, verbatim. Two
commands, \decl{fo_language} and \decl{fo_block}, generate the index types,
summed vocabularies and per-symbol abbreviations that a vocabulary or a
second-order block needs, all of it determined by the arities. Converting the
example of
Appendix~\ref{sec:appendix:example} took its problem module from 314 to 275
lines, the interpretation losing half of its own.

\subsection{\textsf{NP}- and \textsf{coNP}-complete problems}
\label{sec:problems:np}

\begin{table*}
  \caption{Mechanized complexity theory, by what the complexity classes are
    defined over. \emph{Thms} counts completeness theorems on concrete
    problems.
    \textsuperscript{$\ast$}\,\decl{poly-reductions} proves reductions, not
    completeness: the \NP-hardness at the root of its tree rests on an
    admitted lemma (\S\ref{sec:landscape}).}
  \label{tab:comparison}
  \small
  \begin{tabular}{llP{0.155}P{0.215}P{0.115}r}
    \toprule
    \textbf{Library} & \textbf{Prover} & \textbf{Defined over} & \textbf{Classes} & \textbf{Completeness for} & \textbf{Thms} \\
    \midrule
    \decl{coq-library-complexity}~\cite{gaeher2021cooklevin}
      & Rocq         & call-by-value~$\lambda$-calculus & \Ptime, \NP & \NP & 3 \\
    AFP \decl{Cook_Levin}~\cite{balbach2023cooklevin}
      & Isabelle/HOL & multi-tape Turing machines & \textsf{DTIME}, \Ptime, \NP & \NP & 1 \\
    \decl{poly-reductions}~\cite{polyreductions}
      & Isabelle/HOL & \textsf{IMP-} while-language~\cite{kappelmann2025imp} & \Ptime, \NP & none\textsuperscript{$\ast$} & 0 \\
    \decl{Complexitylib}~\cite{complexitylib}
      & Lean         & multi-tape Turing machines & \Logspace, \NL, \Ptime, \NP, \coNP, \PSPACE, \EXPTIME, \NEXPTIME, and 21 others & \NP, \coNP & 3 \\
    Our library
      & Lean         & logic over finite structures & \Logspace, \NL, \Ptime, \NP, \coNP, \DP, \SigmaP{k}, \PiP{k}, \PSPACE, \EXPTIME, \NEXPTIME, \EXPSPACE, \RE, \GI{} degree & all & \dcCompleteThms \\
    \bottomrule
  \end{tabular}
\end{table*}
Figure~\ref{fig:reductions} is the \NP{} part of the catalog, read out of the
elaborated library: every node is a problem proved \NP-complete and every
arrow between problems a proved first-order reduction. Its core is Karp's 21
problems~\cite{karp1972reducibility}, highlighted in the figure and all proved
complete here, \prob{sat}~\cite{cook1971complexity,levin1973universal} among
them.

Hardness enters the graph once, at the arrow out of~\ESO{}. That arrow is the
hardness half of Cook--Levin, proved without a machine:
\decl{sat_hard_of_sigmaSODefinable} sends every problem definable
in~$\Sigma^1_1$ to \prob{sat} -- by Fagin's theorem, every problem in~\NP{} --
and every other problem inherits hardness by composing the reductions along a
path from \prob{sat} to it, with no second-order argument of its own.
Membership is proved separately, and cheaply, by a definability witness per
problem; that is why the graph is not
strongly connected.

Fourteen problems beside Karp's list round the catalog out. Five are
required as intermediate problems;
two are the machine problems of~\S\ref{sec:machines},
\prob{ntm-accept} and one-block \prob{atm-accept},
on the alternating machines of~\citet{chandra1981alternation},
interreducible and complete for~\NP{}; four are common starting points
for hardness proofs elsewhere (\prob{subgraph isomorphism}, \prob{dominating set},
\prob{set splitting}, \prob{edge-weighted steiner tree}); finally, the
last three (\prob{cq evaluation} and \prob{cq
containment}~\cite{chandra1977optimal}, and \prob{graph crawling}) are domain-specific problems the library ships as tutorials.
Appendix~\ref{sec:appendix:example} walks a smaller problem,
\prob{subgraph isomorphism}, through every step of its completeness proof.

Most arrows are the order-free~\fored{}: tags do the work a textbook
first-order reduction gives to the order
(\S\ref{sec:framework:interpretations}), so a gadget that only builds a fixed
number of copies of its input never mentions one. The other forms of
reduction are used when required. First, a problem involving numbers
(e.g., weights) needs its gadget to build them, and a number is a
positional encoding over the universe, so the bit blocks are laid along
the order.
This is why \prob{knapsack} and \prob{partition} are reached by ordered
reductions~\foredord{}. Second, spanning problems, such as \prob{hamilton
circuit}, must visit every vertex, so the meaningless tuples
that other reductions can leave isolated in the output universe would
have to be visited too. Only \prob{vertex cover} to \prob{hamilton circuit}
therefore needs the relativized~\foredrel{}, whose domain formula cuts the
universe down to the real gadget.

The \coNP{} side relies on taking the complement, which yields
completeness results at little cost.
Since \coNP{} is~\ASO, a $\Pi^1_1$-definable problem has a
$\Sigma^1_1$-definable complement, which that same theorem sends to
\prob{sat}; complementing that reduction yields a reduction to \prob{taut}, whence
\decl{taut_hard_of_piSODefinable} and \decl{TAUT_coNP_complete}.

\subsection{The rest of the catalog}
\label{sec:problems:rest}

Problems shown complete for the remaining classes of Table~\ref{tab:classes}
usually fit one of four patterns.
(i)~\emph{The defining logic's own construct, as a problem.} Hardness translates that
logic into the problem's vocabulary: \prob{horn-sat}~\cite{dowling1984linear}
for~\Ptime{}, which is
$\SO\text{-Horn}$, and \prob{det-reach} for~\Logspace{}, which is
$\FO({\leq}, \DTC)$.
(ii)~\emph{The complement of a complete problem}, as \prob{taut} was in
\S\ref{sec:problems:np}. What that costs varies. It is free where the defining
logic is closed under complement, so \prob{qsat}~\cite{stockmeyer1973word},
complete for~\PSPACE{}, is complete for \coPSPACE{} as well; whereas
\prob{unreach} is \NL-complete only through $\NL = \coNL$.
(iii)~\emph{A succinct reading of a problem one exponent down.} Tiling the
$n \times n$ square is in~\NP{}, and tiling the corridor of width~$n$ is
in~\PSPACE{}. Read the instance's elements as addresses, not as
positions and the same two problems, over a grid of side~$2^n$, are
\NEXPTIME-complete and \EXPSPACE-complete.
(iv)~\emph{A classical complete problem}, reached by reduction from the
problem of~(i): \prob{cvp} for~\Ptime{}~\cite{ladner1975circuit} and
\prob{reach} for~\NL{}~\cite{jones1975space}, each complete here under
order-invariant first-order reductions, strengthening the classical
logarithmic-space statements.

\subsection{The \textsf{GI} degree}
\label{sec:problems:gi}

The graph isomorphism problem, \prob{graph-iso}, fits none of the logically
defined classes of Table~\ref{tab:classes}. It is in~\NP, not known to be \NP-complete and not
known to be in~\Ptime{}, and it has no
logical characterization to be complete for. A notion of completeness tied to
a logic would have no place for it at all.

The degree construction of~\S\ref{sec:classes:degrees} gives it one anyway,
since the downward closure of a fixed problem is itself a class. Three
problems are proved complete for the degree of \prob{graph-iso}:
\decl{graphIso_GI_complete}, \decl{digraphIso_GI_complete} and
\decl{dagIso_GI_complete}. Undirected isomorphism is the anchor, that being
what the literature calls~\GI{}, and the other two are interreducible
with it.

Reductions between isomorphism problems are unlike the rest of the catalog,
and the digraph-to-graph gadget runs to some seven hundred and fifty lines for that
reason: a reduction that \emph{translates} an instance is cheap, its
correctness following the construction, whereas one that must \emph{reflect}
isomorphism has to rule out every accidental automorphism of the output, one
argument for each way one could arise.

\begin{toappendix}
\section{Complete Problems per Class}
\label{sec:appendix:completeproblems}

Table~\ref{tab:completeproblems} counts, class by class, the \dcCompleteThms{}
completeness theorems of \S\ref{sec:problems} across \dcCompleteClasses{}
classes. We count theorems, not problems, and the \dcCompleteThms{} theorems
are about \problemSubjects{} distinct problems, a family such as
\prob{qbf}\(_k\) counting once: a class whose complete problem comes in
several forms -- \prob{taut}, \prob{3-dnf-taut} and \prob{3-unsat} for
\coNP{} -- contributes one per form, and a problem complete for two classes
proved equal -- \prob{qsat} for \PSPACE{} and \coPSPACE{} -- contributes a
theorem to each. A theorem that merely restates another -- an
instantiation of a statement already counted for every~$k$, say -- is not
counted twice. The \NP{} row still exceeds the 35 problems of
Figure~\ref{fig:reductions} by three, each a further theorem attached to a
problem already drawn there: \prob{3-colorability} is also proved complete as
\decl{KCol} for every~$k \geq 3$, and \prob{cq evaluation} and \prob{graph
crawling} are also proved complete on their well-formed instances alone
(\S\ref{sec:framework:encodings}).

\begin{table}[t]
  \caption{Completeness theorems proved, by class, with the rows of
    Table~\ref{tab:classes}. \PH{} is the one class of that table with no
    entry, and its empty cell is a theorem, not a gap: a complete
    problem for \PH{} would collapse the hierarchy. \GI{} is set below the
    rule because it is a degree (\S\ref{sec:classes:degrees}).}
  \label{tab:completeproblems}
  \small
  \begin{tabular}{lr}
    \toprule
    \textbf{Class} & \textbf{Completeness theorems} \\
    \midrule
    \completeproblemrows
    \bottomrule
  \end{tabular}
\end{table}

\section{What a Complete Problem Costs}
\label{sec:appendix:problemcosts}
\begin{table*}[htbp]
  \caption{What each problem of the catalog costs: the non-blank source lines
    of its own modules, with the classes it is proved complete for and the
    number of completeness theorems it carries. Not a declaration closure,
    unlike Tables~\ref{tab:cooklevin} and~\ref{tab:machinelevels}.
    The machine models are excluded and costed in
    Table~\ref{tab:machinelevels} instead. Read down the left
    panel, then the right.}
  \label{tab:problemcosts}
  \small
  \setlength{\tabcolsep}{4pt}
  \begin{tabular}{llrr@{\hspace{7em}}llrr}
    \toprule
    \textbf{Module} & \textbf{Complete for} & \textbf{Thms} & \textbf{Lines}
      & \textbf{Module} & \textbf{Complete for} & \textbf{Thms} & \textbf{Lines} \\
    \midrule
    \problemcostrows
    \bottomrule
  \end{tabular}
\end{table*}

We cost the catalog of \S\ref{sec:problems} problem by problem, measuring what
a further problem adds to a library that already has the framework. That
differs from the declaration closures of Tables~\ref{tab:cooklevin}
and~\ref{tab:machinelevels}, which count the framework beneath a theorem, paid
once. So a problem is a module \decl{Problems/X.lean} with its directory
\decl{Problems/X/} -- the vocabulary, the encoding, the definability witness,
the reduction gadgets and the completeness statement -- and we count the
non-blank source lines of those modules. Of the \problemCount{} problems here,
\problemUnderTwoThousand{} come in under two thousand lines. The theorem
counts here sum to those of Table~\ref{tab:completeproblems}, and a problem
whose completeness theorem is proved elsewhere is still costed where its work
is: \decl{REACH_NL_complete} is stated in \decl{TransitiveClosureReach}, but
its subject is defined under \decl{Problems/Reachability/} and is counted
there.

The machine models are the one exception. Their cost is the cost of building a
machine model, the subject of Appendix~\ref{sec:appendix:machinelevels}.
Nothing else is left out. The two wide tilings share a directory with the wide
machines, and are costed here on the modules that are theirs alone, the
machines keeping the rest. So the two tables account for every completeness
theorem between them: \problemTheorems{} of the \dcCompleteThms{} are costed
here, one problem often carrying several, and the remaining
\problemMachineThms{} belong to the machine models, which
Table~\ref{tab:machinelevels} lists, second machines included. Its \coNP{} row
is the one-block instance of the~\PiP{k} theorem above it, and so is not a
further one.
\section{A Worked Example: Subgraph Isomorphism}
\label{sec:appendix:example}

The figures of \S\ref{sec:framework} show the definitions of the
framework. Here we show how a user proves a completeness result against
them. We take one problem of the catalog and walk it through the six
steps every problem follows -- vocabulary, semantics, invariance,
membership, hardness, completeness -- and a seventh, the encoding of
concrete data and its decoding, showing every declaration and eliding
every proof. The problem is \prob{subgraph isomorphism}: does the host
graph contain a subgraph isomorphic to the pattern graph? It is among the
cheapest rows of Table~\ref{tab:problemcosts}, and a fairly
self-contained example. It is also one of the few catalog problems that
come with an encoding and a decoding to concrete Mathlib classes. Of its
474 lines, the problem itself is \decl{Problems/SubgraphIso.lean}, 275
lines, roughly half of them the reduction and its correctness, a
quarter the definability witness, and the rest definitions and
invariance; the remaining 199, \decl{Problems/SubgraphIso/Encoding.lean},
are Step~7.

\subsection*{Step 1: the vocabulary}

An instance carries two graphs, so the vocabulary has two unary marks, telling
the vertices of the pattern from those of the host, and two binary relations,
one adjacency for each. The \decl{fo_language} command declares it, a symbol
and its arity per line, under a prefix for the abbreviations it is to
generate, here \decl{tg} for \decl{twoGraphs}. It generates what one would
otherwise write out: an inductive type of relation symbols indexed by arity,
packaged as a \decl{Language} with no function symbols, whence the
\decl{IsRelational} instance that \decl{DecisionProblem} requires
(\S\ref{sec:framework:problems}), and the four abbreviations \decl{tgPatV},
\decl{tgHostV}, \decl{tgPatE} and~\decl{tgHostE} naming the symbols at their
arity, for use in formulas.

\leanfile{snippets/example-vocabulary.lean}

Nothing forces an element of the universe to be a vertex of either graph.
Elements outside both marks are ``junk elements'' that no condition below mentions. That is
deliberate: it lets an interpretation build such a structure inside a tagged
power of its input universe (\S\ref{sec:framework:interpretations}), as Step~5
does.

\subsection*{Step 2: the semantics}

The property is stated first generically, for arbitrary predicates on a
type, then instantiated to the relations of a structure. \decl{SubgraphIsoOn}
asks for a map that sends pattern vertices to host vertices, injectively on
the pattern, carrying pattern edges to host edges: an injective homomorphism
of the pattern into the host, which is the usual reading (not an induced
subgraph, so that non-edges of the pattern are unconstrained and a clique is
a special case). The map is total on the universe and unconstrained off the
pattern, junk included. \decl{fo_predicates} generates the four shorthands
\decl{TGPatV} to~\decl{TGHostE}, which read the relations of a
\decl{twoGraphs}-structure as predicates, and \decl{HasSubgraphIso} is the
property on structures. Its finiteness conjunct is there for uniformity with
the threshold problems of the catalog, whose cardinality comparisons need
it; by \decl{mem_congr_finite} (\S\ref{sec:framework:problems}) it changes
no complexity-theoretic statement.

\leanfile{snippets/example-semantics.lean}

\subsection*{Step 3: invariance and the bundled problem}

Isomorphism-invariance is the proof obligation of Figure~\ref{fig:problem}.
The generic property transports along any equivalence of types that commutes
with the four predicates -- the map is conjugated by the equivalence, a
twenty-line lemma (\decl{SubgraphIsoOn.of_equiv}, made a biconditional by
\decl{SubgraphIsoOn.equiv_iff}) -- and an isomorphism of structures commutes
with every relation by the library's transport lemmas \decl{relMap_equiv₁} and
\decl{relMap_equiv₂}, so the invariance theorem \decl{hasSubgraphIso_iso} is a
six-line application of the two. Here is its statement, and the bundled
problem:

\leanfile{snippets/example-invariance.lean}

\subsection*{Step 4: membership, by a definability witness}

Membership in \NP{} is \ESO-definability, since that is how the library
defines \NP{} (Table~\ref{tab:classes}), so the witness is a sentence: guess
the map as a binary relation variable, then check it first-order. The
\decl{fo_block} command declares all of it at once. A second-order block is a
finite index type of relation variables with their arities, here the single
variable \decl{map} of arity two. The kernel is a sentence not over
\decl{Language.twoGraphs} but over its sum with the block's own symbols, which
the command names \decl{subgraphSOLang}; and under the second prefix it
generates an abbreviation for every symbol of that sum, the four of Step~1
injected on the left, \decl{sgPatVSym} for \decl{tgPatV} and so on, and the
guessed \decl{sgMapSym} on the right. The prefix says over which vocabulary a
symbol is read: \decl{tg} over \decl{twoGraphs}, as in Step~1, and~\decl{sg}
over \decl{subgraphSOLang}.

\leanfile{snippets/example-block.lean}

The kernel is the conjunction of three clauses, each a first-order sentence
over that sum, and each written in the \lean|fo
\S\ref{sec:problems}: variables are named and bound by \lean|∀| and
\lean|∃|, a relation is applied to terms as \lean|R(x, y)|, the connectives
are the usual ones, and \lean|≐| is equality of terms, kept apart from Lean's
own \lean|=|.
We write $V_p$ and~$V_h$ for the two vertex marks, $E_p$ and~$E_h$ for the two
adjacencies, and~$M$ for the guessed relation.

The first clause says that every pattern vertex is mapped to some host
vertex:
\[
  \forall x_0\, \bigl(V_p(x_0) \rightarrow
    \exists y\, (M(x_0, y) \wedge V_h(y))\bigr).
\]
\leanfile{snippets/example-clause-total.lean}

The second says that the map is injective on the pattern:
\[
  \forall x_0 x_1 x_2\, \bigl(V_p(x_0) \wedge V_p(x_1) \wedge
    M(x_0, x_2) \wedge M(x_1, x_2) \rightarrow x_0 = x_1\bigr).
\]
\leanfile{snippets/example-clause-inj.lean}

The third says that it carries pattern edges to host edges:
\[
  \forall x_0 x_1 x_2 x_3\, \bigl(V_p(x_0) \wedge V_p(x_1) \wedge
    E_p(x_0, x_1) \wedge M(x_0, x_2) \wedge M(x_1, x_3)
    \rightarrow E_h(x_2, x_3)\bigr).
\]
\leanfile{snippets/example-clause-edge.lean}

The kernel is their conjunction. The quantifier~$\exists M$ is no part of it:
the block carries that, so that \decl{SigmaSODefinable 1 SubgraphIso} is the
\ESO{} sentence $\exists M$ of the three clauses.

\leanfile{snippets/example-kernel.lean}

\decl{SigmaSODefinable 1 P} unfolds to a list of one block and a sentence over
the summed vocabulary that agrees with~$P$ on every finite nonempty structure.
Its proof relies on a private lemma
\decl{realize_subgraphKernel} that reads the realization of the kernel back
as a statement of Lean about the guessed relation, some twenty-five lines,
most of them one \decl{simp} call listing the realization lemmas of the
connectives; the two directions are then the mathematics, in thirty lines: an
injective homomorphism is the graph of a function satisfying the three
clauses, and a relation satisfying them yields one by choice.

\subsection*{Step 5: hardness, by a reduction from \prob{clique}}

The reduction is from \prob{clique}, on marked graphs -- a graph with a marked
set whose cardinality is the threshold~$k$. The host is the input graph, and
the pattern is the complete graph on the marked set, so that an injective
homomorphism of the pattern is exactly a clique at least as large as the
marked set. This is the interpretation of
\S\ref{sec:framework:interpretations} with two tags and dimension one,
\decl{Bool} standing for $\{\ell, r\}$: the tag \decl{true} carries the
pattern copy of every vertex and \decl{false} its host copy. The listing reads
against Figure~\ref{fig:interpretation}: for each symbol of the output
vocabulary and each assignment~$t$ of tags to its arguments, a formula over
the input vocabulary of marked graphs, whose mark and adjacency symbols are
\decl{mgMarked} and~\decl{mgAdj}. It is written in the surface syntax of
Step~4, in its \lean|fo
the dimension is one, so an argument is a single element of the input,
\lean|u| the first and~\lean|v| the second, where a hand-written formula would
address them by argument and coordinate as $(0,0)$ and~$(1,0)$. The \lean|if|
chooses between formulas by the tags, and \lean|⊥ᶠ| is the false formula,
apart from Lean's own~\lean|⊥|. The pattern vertices are exactly the pattern
copies of marked vertices, and the host vertices exactly the host copies; two
pattern copies are pattern-adjacent when they are copies of distinct marked
vertices, two host copies host-adjacent when they are copies of distinct
adjacent ones, and every other combination is~$\bot$. The pattern copy of an
unmarked vertex is thus in neither graph, junk the guessed map may send
anywhere, as \S\ref{sec:framework:interpretations} said it would be. Every
formula is quantifier-free, and all but one is a \emph{first-order projection}
in the sense of \S\ref{sec:design:reductions}: a constant, or a single bit of
the input, under a condition on the tags and the coordinates alone. The
pattern-adjacency formula misses only by conjoining the two marks.

\leanfile{snippets/example-reduction.lean}

Correctness is the theorem \decl{hasLargeClique_iff_subgraphIso_map} between
the interpretation and the reduction. Six characterization lemmas, the \decl{clPat_} family, first say what each
relation of the interpreted structure means on tagged copies, each one a
\decl{simp} call through \decl{FOInterpretation.relMap_map}; the two
directions then take fifty lines, an embedding of the marked set into a clique
being turned into an injective homomorphism and back. The threshold of
\prob{clique} is consumed by the shape of the pattern, not by a counting
argument: the one cardinality fact used is that a set at least as large as the
marked set receives an injection from it, \decl{cliqueOn_iff_embedding}, which
\prob{clique} already provides for its own witness. The reduction record then
packages the interpretation with its correctness, exactly as in
Figure~\ref{fig:reduction}.

\subsection*{Step 6: completeness}

Membership is the witness, since \NP{} is \ESO{} by definition; hardness
composes the reduction with the hardness of \prob{clique}, closure under
reductions being part of what a class is (\S\ref{sec:classes:closure});
completeness is the pair.

\leanfile{snippets/example-completeness.lean}

No machine appears anywhere. The hardness of \prob{clique} is that of
\prob{sat}, by the ordered reduction of Figure~\ref{fig:reductions}, and
that of \prob{sat} is the arrow out of~\ESO{} in the same figure.

\subsection*{Step 7: from concrete graphs and back}

The problem so far is stated on structures. A user holds two concrete graphs,
a pattern on~$p$ vertices and a host on~$h$, each a finite set of edges, and
the encoding and decoding of \S\ref{sec:framework:encodings} tie those to
\decl{SubgraphIso}. The concrete instance is a record. Its size is the
textbook one, the vertices and edges of both graphs, and the two bounds below
are stated against it. Its predicate is the textbook one too: an injective map
from the pattern's vertices to the host's carrying every edge to an edge. The
encoder puts the pattern's vertices to the left of a sum and the host's to the
right. The marks then read the side, and the two adjacency relations decide
membership in the two edge sets. The encoder is a plain \decl{def}, so the
compiler vouches that it computes. The bundle of Figure~\ref{fig:encoding}
carries the two bounds as proofs, elided below. The universe is the vertex
set, so nothing is padded, and an edge set has at most quadratically many
elements, so nothing is compressed. Faithfulness closes the listing. Its
proof, and the decoder's, go through one lemma: against an enumeration of the
pattern vertices and one of the host vertices, the generic property of Step~2
is the textbook predicate (\decl{subgraphIsoOn_iff_concrete}); the encoding
enumerates them by the two injections of the sum.

\leanfile{snippets/example-encoding.lean}

The decoding direction reads hardness back (\S\ref{sec:framework:encodings}),
and here it needs no well-formedness condition. The semantics ignores the
elements in neither mark, and never relates the pattern role and the host role
of one element, so a structure marking an element as both is the same instance
as one where that element is split in two. The decoder therefore reads every
presented structure back, a \decl{FinPresentation} being a finite structure
whose relations come as computable tables: the pattern vertices are the
pattern-marked elements
and the host vertices the host-marked ones, each enumerated in order
(\decl{Finset.orderEmbOfFin}), and the edges are read off the tables. Its
soundness is the lemma above with the other pair of enumerations, and its
totality is \decl{rfl}, the condition~$\Wtt$ of
\S\ref{sec:framework:encodings} being~$\top$. The consequence is the last
statement: every finite nonempty structure is decided by \decl{SubgraphIso}
exactly as some pair of concrete graphs is by the textbook predicate, so the
hardness of Step~5 is about concrete graphs and not about junk.

\leanfile{snippets/example-decoding.lean}

The compiler enforces that the encoder and the decoder compute: it would
demand \decl{noncomputable} of a definition deciding an undecidable predicate.
Their complexity is not bounded, the interface measuring nothing against a
machine, and the reader sees from \decl{sgRelBool} that it decides membership
in two finite sets.

\subsection*{What the other problems pay for}

The more expensive rows of Table~\ref{tab:problemcosts} pay for three things
this problem does not need. An ordered reduction, when the gadget needs an
element the input does not single out -- the fresh variable of the reduction
to \prob{nae-sat}, the three marked copies of the minimum in the one to
\prob{chromatic number}. A number laid along the order, when the target
carries weights (\S\ref{sec:problems:np}). And a well-formedness condition,
when a universe has junk that no decoder can read back without deciding
something: the two tutorials of the library, \prob{cq evaluation} and
\prob{graph crawling}, restrict their completeness theorems to the well-formed
instances -- a website with exactly one root, for the latter -- where Step~7
restricts nothing.

\end{toappendix}

\begin{table*}
  \caption{Lines strictly necessary to prove the Cook--Levin
    theorem (\NP-completeness of \prob{sat} under each library's machine model).
    \emph{Closure} counts the declarations the proof reaches; \emph{In full}
    counts the same modules in full; \emph{External libraries} is what the closure reaches in the
    libraries the development imports.}
  \label{tab:cooklevin}
  \small
  \setlength{\tabcolsep}{4pt}
  \begin{tabular}{llrrrl}
    \toprule
    \textbf{Library} & \textbf{Theorem measured} & \textbf{Modules} & \textbf{Closure} & \textbf{In full} & \textbf{External libraries} \\
    \midrule
    \decl{coq-library-complexity}~\cite{gaeher2021cooklevin}
      & \decl{CookLevin}
      & \gkModules & \gkOwnLines & \gkModuleLines
      & \gkExtLines\ (undec.\ lib.\ + \textsf{Coq} stdlib) \\
    AFP \decl{Cook_Levin}~\cite{balbach2023cooklevin}
      & \decl{NP_complete_SAT}
      & \balbachModules & \balbachOwnLines & \balbachModuleLines
      & \balbachExtLines\ (\textsf{HOL} only) \\
    \decl{Complexitylib}~\cite{complexitylib}
      & \decl{NPComplete_language}
      & \clSatModules & \clSatOwnLines & \clSatModuleLines
      & \clSatExtLines\ (Mathlib, core) \\
    Our library
      & \decl{SAT_complete_for_ntmAccept}
      & \dcBothModules & \dcBothOwnLines & \dcBothModuleLines
      & \dcBothExtLines\ (Mathlib, core) \\
    \bottomrule
  \end{tabular}
\end{table*}

\section{Related Work}
\label{sec:related}
\label{sec:landscape}
\label{sec:related:cooklevin}
\label{sec:related:reductions}

Lean's general-purpose libraries stop short of complexity classes. Mathlib's
\decl{Computability} tree offers polynomial-time computability of functions
and the many-one reducibility of computability theory; the official Lean
computer science library~\cite{cslib} adds Turing machines with
\decl{TimeComputable} and \decl{PolyTimeComputable} predicates. Neither
defines a complexity class, a reduction carrying a resource bound, or a
completeness result.

Four developments do define complexity classes, as recorded in
Table~\ref{tab:comparison}. \citet{gaeher2021cooklevin} work over the
call-by-value~$\lambda$-calculus, their three completeness theorems all in
the module proving Cook--Levin; \mbox{\citet{balbach2023cooklevin}} works
over multi-tape Turing machines, Cook--Levin being the completeness theorem
proved.
\decl{poly-reductions}~\cite{polyreductions} proves reductions rather than
completeness: 26~problems, all 21~of Karp's among them, form a
tree rooted at~\prob{sat}, whose \NP-hardness is proved only inside an
Isabelle locale assuming the step from a polynomial-time verifier to a
reduction to~\prob{sat}, a lemma left admitted.
\decl{Complexitylib}~\cite{complexitylib}, back over multi-tape machines,
has the widest inventory: 29~classes at the revision we cite, with
completeness for \NP{} and \coNP{} on three concrete problems, and, beside
them, first- and second-order syntax over finite structures with
one-dimensional first-order reductions, Fagin's theorem being open on its
roadmap and no class defined through them.
Machine models have a clock in return, so \decl{Complexitylib} proves the
deterministic time hierarchy theorem and circuit lower bounds, and
\citet{forster2020turing} verify a universal machine and a multi-tape to
single-tape compiler with polynomial overhead, results our library cannot
state.

These four are the only developments we could find; others advertising
machine-checked complexity results were excluded once we confirmed their
headline theorems admitted or vacuous, reducing to \decl{True} or resting on
classes that are opaque axioms.

Cook--Levin is the one theorem all but \decl{poly-reductions} prove, and
Table~\ref{tab:cooklevin} measures, for each, the lines of code
\emph{strictly necessary} to prove it: the dependency closure of the
theorem, not of the files holding it. Every row measures
the machine form of the theorem, \prob{sat} complete for the \NP{} that the
development's own machine model defines under a polynomial time bound. We
report each closure split into the development's own modules and the external
libraries beneath them, a small count resting on a large library being a
different achievement, and give the same modules counted in full beside it.
How each closure is computed is Appendix~\ref{sec:appendix:closures}.

\begin{toappendix}
\section{How the Closures of Table~\ref{tab:cooklevin} Are Computed}
\label{sec:appendix:closures}

In Lean, proofs are terms, so we take the declarations a theorem's proof term
mentions, close under that relation through statements and proof terms alike,
map each declaration to its source range, merge overlapping ranges per module,
and count the lines of code they cover. A line counts when it carries
something other than whitespace, comment, or docstring, a convention reused in
every measurement in this paper. The Rocq and Isabelle rows are the same
measurement made with each prover's own machinery: a kernel-level dependency
graph from \decl{coq-dpdgraph}, and a walk of the proof body with Isabelle's
\decl{Proofterm.fold_body_thms}.
\end{toappendix}

We checked what each theorem relies on with each prover's own facility: the
Rocq row depends on no axiom, the
Isabelle row on no oracle, and the two Lean rows on the three standard
axioms \decl{propext},
\decl{Classical.choice}, and \decl{Quot.sound}; no row is admitted. Over the
whole libraries rather than these closures, ours declares no axiom of its own
and contains no \decl{sorry}.

The closures differ by a factor of about seven on the developments' own side,
from \dcBothOwnLines{} lines to \balbachOwnLines{}; the two Lean rows compare
most directly, using the same prover and the same libraries beneath, at
\clSatOwnLines{} lines against \dcBothOwnLines{}. Line counts across four
provers are a weak measure at best: the scopes differ, the inputs differ --
binary strings in the Isabelle and \decl{Complexitylib} rows, encoded
$\lambda$-terms in the Rocq one, finite structures here
(\S\ref{sec:design:structures}) -- and so do the libraries beneath them.
Read as an order of magnitude, though, they say
something: ours is the smallest of the four, and that with the whole machine
bridge of \S\ref{sec:machines} inside the closure
(Appendix~\ref{sec:appendix:machinelevels} costs it class by class). That is
evidence that descriptive complexity is a useful route to complexity
results.

The Rocq library of undecidable
problems~\cite{forster2020undecidability} fixes one notion of reduction,
grows a catalog around it, and takes contributions from others, which is the
shape of our library one level down, with first-order reductions in place of
many-one ones and completeness in place of undecidability;
\citet{grange2024cookbook} specify reductions as graphical recipes
equivalent to quantifier-free first-order interpretations, so that candidate
reductions can be validated automatically. On the model-theoretic side, \citet{kirst2022trakhtenbrot} prove
Trakhtenbrot's theorem in Rocq, \prob{finsat} being the problem closest to it
in our catalog; first-order logic over finite relations has been mechanized
for query evaluation~\cite{raszyk2022eval}, and the least-fixpoint semantics
of Datalog, the iteration behind Gr{\"a}del's theorem in
\S\ref{sec:classes:relations}, more than
once~\cite{benzaken2017datalog,tantow2025datalog}; and the logic--automata
connection has verified decision procedures for
WS1S~\cite{traytel2015ws1s,doczkal2018reglang}. We know of no prior
mechanization of Ehrenfeucht--Fra{\"i}ss{\'e} games, of the fixpoint and
transitive-closure logics, or of the capture theorems of
\S\ref{sec:classes}.

\section{Design Decisions and Limitations}
\label{sec:design}

Two decisions shape the library: hardness is proved by first-order reduction,
and an instance is a finite structure. We take each in turn, then the
questions the approach cannot reach.

\subsection{First-order vs.\ Karp reductions}
\label{sec:design:reductions}

Hardness is classically defined by Karp reductions, that is, many-one
reductions computable in polynomial time, or by their logarithmic-space
refinement (\S\ref{sec:background:red}). Ours are first-order, and every
first-order reduction is both, so hardness proved under~\fored{} transfers
verbatim to the classical notion.

The converse fails already against logarithmic space, and the library proves
it (\decl{exists_dtcReduction_not_orderedReduction}). In $\FO({\leq}, \DTC)$,
\prob{even} of \S\ref{sec:classes:lower} reduces to \prob{nonempty-mark},
whether an element is marked: mark everything when the universe is even,
nothing when it is odd, one walk along the order. No first-order reduction
does so, the target being first-order definable and \prob{even} not. That walk
never has a choice, so this is the deterministic notion, the many-one
reduction of the textbooks: completeness under the weaker reduction is the
stronger statement.

In practice the restriction has cost the library nothing. Every reduction in
the catalog is $\FO({\leq})$-definable, across all \dcCompleteThms{}
completeness results and \dcCompleteClasses{} classes (\S\ref{sec:problems});
no problem had to be dropped, and none needed a weaker statement, because its
textbook reduction would not fit in the logic. Where the textbook route does
resist, another is available. \citet{murray2017nonnphardness} observe that
reductions with local structure suffice for almost all completeness results,
and name one potential counterexample: the usual reduction from \prob{subset-sum} to
\prob{partition} outputs two numbers that require summing every weight in the
instance, so it does not seem computable even by \ACz{} circuits of
size~$2^{n^{o(1)}}$. The library reaches \prob{partition} by an ordered
first-order reduction from \prob{nae-sat}~\cite{schaefer1978complexity}
instead, and a recent note~\cite{risco2025partition} gives first-order
projections.

Such reductions are usually first-order
\emph{projections}~\cite{immerman1987languages}, refining those
of~\citet{skyum1985complexity}: each bit of the output is a constant, or one
bit of the input, or its negation, independently of the input. Natural
complete problems seem to stay complete under
them~\cite{allender1997first}, although an artificial set complete for~\NP{}
under $\ACz[\mathrm{mod}\;2]$ reductions but not under \ACz{} ones
exists~\cite{agrawal1997reducing}; what our catalog adds is machine-checked
evidence for the problems one actually meets.

\subsection{Finite structures vs.\ strings}
\label{sec:design:structures}

A machine model is classically defined on strings, with its resource bounds
given as explicit functions of the input length. Here an instance is a finite
structure, a machine is data inside it, and the budget is a count of universe
elements, the polynomial coming from the dimension of the reduction that built
the instance (\S\ref{sec:machines:problem}); no arithmetic appears in the
vocabulary.

The two conventions differ most in how order enters. Strings come ordered and
structures do not, so order-invariance is an explicit variant here,
\foredord{}, and theorems whose whole content is the \emph{absence} of an
order become statable: Abiteboul--Vianu (\S\ref{sec:classes:relations}), and
the failure of order-free \FO(\IFP) to capture \Ptime{}
(\S\ref{sec:classes:lower}), which is why the~$\leq$ carried by the fragments
of Table~\ref{tab:classes} is necessary. Neither convention dominates, though:
the string one makes cost functions and hierarchy theorems natural, and those
are exactly what we give up (\S\ref{sec:design:limits}).

The two meet at the top of the catalog. Finite structures over a finite
vocabulary carry a numbering, so a problem denotes a set of codes, and
\decl{mem_RE_iff_rePred} proves \ESOnew-definability to coincide with
Mathlib's own \decl{REPred} on that set. The witness is \prob{codehalt}, where
a partial recursive code \emph{is} an instance, its syntax tree flattened, so
nothing is simulated. Below \RE{} the agreement with the usual presentations
over string encodings is
classical~\cite{fagin1974generalized,immerman1999descriptive} and is not
formalized here.

The encodings that carry a result back to concrete data
(\S\ref{sec:framework:encodings}) raise the same question. They require
polynomial size bounds in both directions and a faithfulness proof, but the
encoding itself may be any computation, so a completeness statement
transported through one inherits a cost that is never measured. A machine
model faces the same step: its inputs are strings, so a graph or a formula has
to be written out as one before the machine runs, and that cost goes
unmeasured too.

\subsection{What the descriptive approach cannot do}
\label{sec:design:limits}

There is no cost model here, hence no $O(\cdot)$, no fine-grained complexity
and no~$\Delta^p_k$. It is therefore impossible to prove a hierarchy theorem:
separating \textsf{DTIME}$(f)$ from \textsf{DTIME}$(2^f)$ is a statement about
a clock, and \decl{Complexitylib} proves the deterministic time hierarchy
theorem because it carries one (\S\ref{sec:landscape}), whereas here there is
nothing to diagonalize against.

\section{Conclusion}
\label{sec:conclusion}

Taking a decision problem to be an isomorphism-invariant property of finite
structures, a complexity class to be a definability predicate, and hardness to
be a first-order reduction, we obtain a library in which membership and
completeness are logical statements, the relations between classes are proved
inside the logic, and the machine models are theorems.
New results come cheaply on that footing: membership is a sentence, hardness a
reduction, completeness the pair of them, and a problem joins the catalog for a
median of \problemMedian{} lines of its own modules -- which is how
\dcCompleteThms{} completeness theorems across \dcCompleteClasses{} classes
were reached.

Three extensions look worthwhile. Locality, in Hanf's and Gaifman's
forms~\cite{libkin2004elements}, would give the inexpressibility results that
now need a strategy built by hand. Counting fits the same idiom: the
quantitative second-order logic of~\citet{arenas2020descriptive} layers
arithmetic over an unchanged Boolean one, so that $\#\mathsf{P}$ arrives as
$\Sigma\mathrm{QSO}(\FO)$ with \#\prob{sat} complete for it. And tracking
quantifier-freeness through composition would carry the catalog to the
projections of~\S\ref{sec:design:reductions}.

\ifarxiv
The library is available at \dcRepo{} under the Apache 2.0 license; the
numbers reported here were measured at release 1.2.2.
\else
The library will be released as open-source under the Apache 2.0 license once
anonymity requirements are lifted.
\fi

\begin{acks}
  This work was funded in part by the French government under management
  of Agence Nationale de la Recherche (ANR) as part of the ``France
  2030''
  program, reference ANR-23-IACL-0008 (PR[AI]RIE-PSAI). It is also part
  of the program DesCartes and is supported by the National Research
  Foundation, Prime Minister's Office, Singapore under its Campus for
  Research Excellence and Technological Enterprise (CREATE) program. We
  are grateful to Dan Suciu who suggested FO reductions as the right tool
  to formalize reductions in Lean.

  The Lean code of the library, including
  most proofs and docstrings, was written with the assistance of generative
  models from Anthropic (Claude); the formalization choices, the library
  architecture, the main definitions and theorems, and their alignment with
  the standard ones were designed and written or reviewed by the authors,
  who take responsibility for the whole.
\end{acks}

\ifarxiv\else\clearpage\fi

\ifarxiv\else\balance\fi
\bibliographystyle{ACM-Reference-Format}
\bibliography{references}

\end{document}